\documentstyle[12pt]{article}
\newcommand{\nsect}{\setcounter{equation}{0}
\def\theequation{\thesection.\arabic{equation}}\section}
\newcommand{\nappend}{\setcounter{equation}{0}
\def\theequation{\rm{A}.\arabic{equation}}\section*}

\catcode`\@=11

\def\marginnote#1{}
\newcount\hour
\newcount\minute
\newtoks\amorpm
\hour=\time\divide\hour by60
\minute=\time{\multiply\hour by60 \global\advance\minute by-\hour}
\edef\standardtime{{\ifnum\hour<12 \global\amorpm={am}%
        \else\global\amorpm={pm}\advance\hour by-12 \fi
        \ifnum\hour=0 \hour=12 \fi
        \number\hour:\ifnum\minute<10 0\fi\number\minute\the\amorpm}}
\edef\militarytime{\number\hour:\ifnum\minute<10 0\fi\number\minute}
\def\draftlabel#1{{\@bsphack\if@filesw {\let\thepage\relax
   \xdef\@gtempa{\write\@auxout{\string
      \newlabel{#1}{{\@currentlabel}{\thepage}}}}}\@gtempa
   \if@nobreak \ifvmode\nobreak\fi\fi\fi\@esphack}
        \gdef\@eqnlabel{#1}}
\def\@eqnlabel{}
\def\@vacuum{}
\def\draftmarginnote#1{\marginpar{\raggedright\scriptsize\tt#1}}
\def\draft{\oddsidemargin -.5truein
        \def\@oddfoot{\sl preliminary draft \hfil
        \rm\thepage\hfil\sl\today\quad\militarytime}
        \let\@evenfoot\@oddfoot \overfullrule 3pt
        \let\label=\draftlabel
        \let\marginnote=\draftmarginnote
   \def\@eqnnum{(\theequation)\rlap{\kern\marginparsep\tt\@eqnlabel}%
\global\let\@eqnlabel\@vacuum}  }

\def\preprint{\twocolumn\sloppy\flushbottom\parindent 1em
        \leftmargini 2em\leftmarginv .5em\leftmarginvi .5em
        \oddsidemargin -.5in    \evensidemargin -.5in
        \columnsep 15mm \footheight 0pt
        \textwidth 250mmin      \topmargin  -.4in
        \headheight 12pt \topskip .4in
        \textheight 175mm
        \footskip 0pt
        \def\@oddhead{\thepage\hfil\addtocounter{page}{1}\thepage}
        \let\@evenhead\@oddhead \def\@oddfoot{} \def\@evenfoot{} }

\def\titlepage{\@restonecolfalse\if@twocolumn\@restonecoltrue\onecolumn
     \else \newpage \fi \thispagestyle{empty}\c@page\z@
        \def\thefootnote{\fnsymbol{footnote}} }

\def\endtitlepage{\if@restonecol\twocolumn \else  \fi
        \def\thefootnote{\arabic{footnote}}
        \setcounter{footnote}{0}}  

\catcode`@=12
\relax

\def\bea{\begin{array}}
\def\bem{\begin{displaymath}}
\def\beq{\begin{equation}}

\def\eea{\end{array}}
\def\eem{\end{displaymath}}
\def\eeq{\end{equation}}

\def\Im{\mathop{\rm Im}}

\def\NP#1#2#3{Nucl. Phys. \underline{#1} (19#2) #3}
\def\ov{\overline}

\def\PL#1#2#3{Phys. Lett. \underline{#1} (19#2) #3}
\def\PR#1#2#3{Phys. Rev. \underline{#1} (19#2) #3}
\def\PRL#1#2#3{Phys. Rev. Lett. \underline{#1} (19#2) #3}
\def\Re{\mathop{\rm Re}}

\def\s2w{\sin^2 \theta_W}
\def\Tr{\mathop{\rm Tr}}
\def\und{\underline}
 \let\vev\VEV

\def\dalpha{{\dot\alpha}}

\def\crbig{\\\noalign{\vspace {3mm}}}
\def\bigint{{\displaystyle\int}}
\def\S{\Sigma}
\def\G{\Gamma}
\def\L{{\cal L}}
\def\SG{S_{\Gamma}}
\def\comp{(z_0\ov z_0)}
\def\S{\Sigma}
\def\G{\Gamma}
\def\L{{\cal L}}
\def\SG{S_{\Gamma}}

\def\c{\cal}

\relax

\begin{document}
\topmargin-2.4cm
%
%
%
%
\begin{titlepage}
\begin{flushright}
NEIP--93--007 \\
IEM--FT--83/94 \\
hep--th/9402007 \\
January 1994
\end{flushright}
\vskip 0.2in
\begin{center}{\Large\bf
The Linear Multiplet and Quantum\\
\vskip .1in
Four-Dimensional String Effective Actions$^*$}
\vskip .2in
{\bf Jean-Pierre Derendinger, Fernando Quevedo}
\vskip .1truecm
{\it Institut de Physique \\
Universit\'e de Neuch\^atel \\
CH--2000 Neuch\^atel, Switzerland}
\vskip.3truecm
and
\vskip.3truecm
{\bf Mariano Quir\'os}
\vskip.1truecm
{\it Instituto de Estructura de la Materia \\
CSIC Serrano 123 \\
E--28006 Madrid, Spain}
\end{center}
\vskip .2in
\begin{center}
{\bf Abstract}
\end{center}
\begin{quote}
In four-dimensional heterotic superstrings, the dilaton and
antisymmetric tensor
fields belong to a linear $N=1$ supersymmetric multiplet $L$.
We study
the lagrangian describing the coupling of one linear multiplet to
chiral and gauge multiplets in global and local supersymmetry, with
particular emphasis on string tree-level and loop-corrected
effective actions.
This theory
is dual to an equivalent one with chiral
multiplets only.
But the formulation with a linear multiplet appears
to have decisive advantages beyond string tree-level
since, in particular, $\langle L \rangle$ is the string
loop-counting parameter and the duality transformation
is in general not exactly solvable beyond tree-level.
This formulation allows us to easily
deduce some powerful
non-renormalization theorems in the effective theory
and to obtain explicitly some loop
corrections to the string effective supergravity for simple
compactifications.
Finally,
we discuss the issue of supersymmetry breaking by gaugino
condensation using this formalism.
\end{quote}
\vskip.1truecm
\hrule width 4.5cm
\vskip.1truecm
$^*$Work supported in part by the Swiss National Foundation,
the European Union (contracts SC1*--CT92--0789 and CHRX--CT92--0004)
and the CICYT (contract AEN90--0139).

\end{titlepage}
\setcounter{footnote}{0}
\newpage
%
%

\nsect{Introduction}

The construction of four-dimensional effective actions for superstring
compactifications is of fundamental importance for  the
study of the phenomenological implications of the theory.
A substantial amount of
work  has been performed in this respect during the past few years.
Different techniques have been successfully applied to compute string
tree-level lagrangians for
many classes of string compactifications. But the extension of
these techniques
to higher orders in string perturbation theory is still a challenge.
At present, only the one-loop corrections to gauge coupling
constants have been explicitly computed in some simple classes
of superstring theories [\ref{M}, \ref{K}, \ref{DKL},
\ref{ANT}], whereas the corrections to the rest of the
lagrangian are simply unknown.
We present in this article a general discussion  of effective lagrangians
describing the couplings of one linear supersymmetric multiplet
[\ref{linear}, \ref{S}, \ref{Betal}, \ref{Oetal}] to chiral
multiplets in $N=1$ global and local supersymmetry in four-dimensions.
The motivation for this is the observation [\ref{DFKZ1}]
that the origin and the
structure of the string one-loop corrections to the effective
gauge coupling constants are particularly easy to understand
using the linear multiplet. It is a natural approach to
the effective supergravity of superstrings to consider
lagrangians with a linear multiplet [\ref{CFV}, \ref{BGG}, \ref{ABGG}],
and we will argue that this formalism is actually the most
convenient  for describing low-energy couplings beyond tree-level.

The importance of the linear multiplet for the superstring
effective action is due to the fact that the gravity sector of
superstrings, which is universal, contains an antisymmetric
tensor $b_{\mu\nu}$ and a real scalar, the dilaton, along
with the
Majorana spinor partner. This is precisely the particle content
of the linear multiplet. There is a gauge symmetry related to
$b_{\mu\nu}$,
$$
b_{\mu\nu} \longrightarrow b_{\mu\nu} + \partial_\mu b_\nu -
\partial_\nu b_\mu,
$$
where $b_\nu$ is arbitrary.
We can immediately remark that this symmetry does not provide
any constraint on the couplings of the linear multiplet $L$. It is
a consequence of the mere existence of $L$. The
linear multiplet can always be transformed into a chiral
multiplet $S$, the antisymmetric tensor being equivalent to a
pseudoscalar. But the dual theory has a different symmetry content.
The chiral multiplet only appears in the combination $S+\ov S$
in the superfield formulation of the lagrangian.
This means that the dual
theory has an invariance under a Peccei--Quinn
symmetry which shifts the pseudoscalar component $\Im s$ of $S$.
The existence of symmetries acting on $b_{\mu\nu}$ and $\Im s$
is at the origin of the duality
transformation and its inverse.

In view of the equivalence of linear and chiral multiplets through
duality transformations, it could seem useless to work out effective
lagrangians using a linear multiplet since the lagrangians for
chiral multiplets coupled to supergravity are well understood whereas
there is not a well-developed formalism
for describing the general couplings to
a linear multiplet. This is essentially true for the tree-level
string effective actions for which the duality transformation can be
exactly solved to give the
explicit form of the effective action in terms of chiral fields
only [\ref{W}]. However, one should notice
that the duality transformation, which is a generalized Legendre
transformation, cannot always be performed analytically at the
higher-loop level.
A perturbative
treatment will in general lose some information and obscure
the symmetry content of the theory.
This is the situation encountered in ref. [\ref{DFKZ2}], where an
effective lagrangian could be constructed with the linear multiplet $L$,
the dual theory using a chiral multiplet $S$ being known only
perturbatively.

 Moreover, it appears that the duality transformation
does not respect string perturbation theory, which is used to obtain
the contributions to the effective theory. Suppose one generates the
effective theory as a formal series ${\cal L}_{eff.} =
{\cal L}_0 + \sum_{n\ge1}^N\Delta{\cal L}_n$, $n$
indicating the string loop
order. At a fixed order $N$, the duality transformation applied
to ${\cal L}_{eff.}$ will severely
mix the various contributions, with the consequence that string
perturbative expansion becomes hard to identify in the dual theory.
The main reason for this is that $\langle L\rangle$
is the string loop-expansion parameter, whereas in the dual theory
the field $S$ is defined order by order in perturbation theory
and its relation to the loop-counting parameter in the
dual theory is not clear. Furthermore, at a given $N$,
the string loop amplitudes have a clear interpretation in
terms of the lagrangian for $L$ but not  for the
dual theory.
An explicit example of this phenomenon already exists.
 String one-loop threshold
corrections to gauge coupling constants in symmetric $(2,2)$
orbifolds, as computed in refs. [\ref{DKL}, \ref{ANT}], can be easily
and naturally interpreted using the linear multiplet [\ref{DFKZ1}]
as corrections to the gauge couplings in the effective lagrangian.
In the dual theory however, these one-loop contributions appear
in the K\"ahler potential, and not in gauge kinetic terms.

These remarks suggest that since the natural partner of the graviton
in the massless sector of heterotic superstrings is an antisymmetric
tensor, string information would have a more
natural translation in an effective supergravity with a linear
multiplet, at least when string loop contributions are considered.
This is the point of view that we will adopt in this
paper.

In general, supersymmetric theories with linear multiplets are more
constrained than theories formulated with chiral multiplets only.
For instance, the superpotentials for the $S$ field  which have been
used in the study of supersymmetry breaking in four-dimensional
superstrings [\ref{DRSW}] are
not compatible in general with the duality transformation
\footnote{Previous attempts
[\ref{FILQ}] have been made to make these
superpotentials consistent with another duality symmetry,
target-space duality [\ref{duality}], which should not be confused with the
duality symmetry we are referring to in the present article.}.
This fact raises important questions since the origin of the $S$
superfield is precisely the presence of the
antisymmetric tensor in strings.
We will therefore address the issue of the formulation of
dynamical supersymmetry breaking by gaugino condensates
in the linear multiplet formalism, a treatment which guarantees
the existence of versions equivalent by duality.

The organization of the present article is as follows. In the next section,
we present a concise description of global $N=1$ supersymmetric actions with
one linear superfield coupled to chiral matter. This section is a
preparation to the more relevant case of supergravity.
We concentrate mainly on the issues of interest
in the study of the effective potential,
supersymmetry breaking and gaugino condensation.
We discuss in detail the supersymmetric duality
transformation [\ref{S}], which maps this theory
to one with only chiral superfields,
including explicit expressions for each of the components of the
 chiral multiplet $S$ in terms of those of $L$.
The comparison of the two dual versions reveals that the interpretation
of the nature of supersymmetry breaking by gaugino condensation is somewhat
ambiguous in these models. In the formulation with a linear multiplet,
insertion of expectation values of gaugino bilinears is clearly an explicit
breaking of supersymmetry. An intuitive reason is that the linear multiplet
does not possess any auxiliary field.
On the other hand, in the dual theory in terms of $S$, such insertions
generate a somewhat degenerate form of spontaneous breaking.
These features of the globally supersymmetric case depend
mainly on the auxiliary field structure
of the theory. They have then a straightforward extension to local
supersymmetry.

Sections 3 and 4 describe the relevant aspects of supergravity theories
with a linear multiplet.  We use the superconformal approach to
minimal supergravity, as reviewed
in [\ref{KU}]. In section 3, we present the basics of our
techniques and
 give a detailed discussion of
the gauge fixing
of scale invariance in a superconformal theory with a linear multiplet.
The superconformal approach followed here is  convenient  for
several reasons. We choose to fix conformal invariance by
imposing a canonical Einstein term, but in this approach non-canonical
choices are equally possible, such as the one corresponding to the
string $\sigma$-model metric, which may be eventually more appropriate.
 It is also convenient for the discussion
of the renormalization-group behaviour of parameters which will
be used in section 5. Section 4 is devoted to the calculation
of these supergravity lagrangians in components. This is a
more complicated task than the corresponding calculations in
the chiral superfield case [\ref{CFGVP}].
The main technical complication resides in solving the conformal
gauge fixing equation giving the canonical Einstein lagrangian.
We however can explicitly write the scalar potential and
kinetic terms for a general class of K\"ahler invariant models.
We reproduce the standard expressions of ref. [\ref{CFGVP}] as a particular
case when the linear multiplet decouples.

In section 5, we start the discussion of the string theory case.
First we present the string tree-level lagrangian
in terms of the linear multiplet and discuss some
powerful non-renormalization
theorems that can be obtained for this lagrangian using simple
properties of the linear multiplet. In particular
the superpotential and the K\"ahler potential, being only functions
of the chiral fields, do not get renormalized as long
as this approach is valid. Also we emphasize that knowledge of
the loop corrections of the gauge coupling constant is
almost sufficient to
obtain the corresponding correction to the full effective action.
We then consider the one-loop corrections to a  class of string models
following ref. [\ref{DFKZ2}]. For simplicity, we  restrict ourselves
to an $E_8$ hidden sector without
matter and consider the couplings of the linear superfield $L$ and
the chiral superfields of the theory, including an overall modulus
field $T$. The (Wilson) effective action in terms of
 superfields is obtained from
the one-loop corrections to the gauge coupling constants as
computed in [\ref{DKL}]. The loop corrections
can be interpreted [\ref{DFKZ1}] as
the Green--Schwarz [\ref{GS}] counterterms cancelling the K\"ahler
sigma-model anomalies associated with target-space duality transformations.
This mechanism is dictated by the fact that target-space duality
is an exact, quantum symmetry of the superstring.
A natural extension of this treatment of anomalies in the
effective field theory is to consider conformal anomalies
in the context of superconformal supergravity,
as suggested in [\ref{DFKZ2}]. In this approach, we
rederive from these
actions the field-dependent renormalization-group
equations and, from them,
obtain the field-dependent renormalization group invariant scale which
characterizes the strength of gauge forces in the hidden sector
and is useful
in the discussion of gaugino condensation and supersymmetry breaking.

Section 6 is devoted to the explicit calculation, in components,
of the loop-corrected
effective actions of the models discussed in section 5.
We present  the scalar kinetic terms and
the scalar potential including gaugino bilinears.
This allows us to make a short discussion of supersymmetry breaking
by gaugino condensation in this formalism. We mention
briefly how the different approaches used in the chiral
case could be implemented with the linear multiplet.
In particular, we remark the incompatibility with the
linear multiplet formalism of the introduction of nontrivial
superpotentials for the $S$ field as a result of gaugino
condensation. We close with some remarks on
the main results of this paper.

\nsect{The linear multiplet and global supersymmetry breaking}

Even if we will essentially be interested in  supergravity effective
theories involving a linear multiplet,
some aspects can be discussed in the simpler context of global
supersymmetry. This is the purpose of this section, which will
also present the main features of theories with a linear multiplet.

\subsection{Sigma-models with a linear multiplet}

In global $N=1$ supersymmetry, the real linear multiplet is a real
vector superfield such that its second supersymmetry covariant
derivatives vanish:
\beq\label{linconst}
{\cal DD}\, L = \ov{\cal DD}\, L =0.
\eeq
The supersymmetry covariant derivatives are\footnote{
We use conventions similar to Wess and Bagger [\ref{WB}], with
signature $(-,+,+,+)$ for the space-time metric.}
\beq\label{covder}
{\cal D}_\alpha = {\partial\over\partial\theta^\alpha} +
i(\sigma^\mu \ov\theta)_\alpha\partial_\mu, \qquad
\ov {\cal D}_\dalpha = -{\partial\over\partial\ov\theta^\dalpha}
- i(\theta\sigma^\mu) _\dalpha \partial_\mu,
\eeq
and ${\cal DD}= {\cal D}^\alpha {\cal D}_\alpha = -\epsilon^
{\alpha\beta}{\cal D}_\alpha {\cal D}_\beta$,
$\ov{\cal DD}= \ov {\cal D}_\dalpha \ov {\cal D}^\dalpha$.
Solving the constraints leads to a component expansion of the
form
\beq\label{linexp}
L= C + i\theta\chi - i\ov\theta\ov\chi + \theta\sigma^\mu\ov\theta\,
v_\mu -{1\over2} \theta\theta\ov\theta (\partial_\mu\chi\sigma^\mu)
-{1\over2} \ov{\theta\theta}\theta (\sigma^\mu\partial_\mu\ov\chi)
-{1\over4} \theta\theta\ov{\theta\theta}\, \Box C.
\eeq
Furthermore, the vector field $v_\mu$ has vanishing divergence,
\beq\label{Vconst}
\partial_\mu v^\mu = 0,
\eeq
hence
\beq\label{bmunu}
v_\mu = {1\over\sqrt2}\epsilon_{\mu\nu\rho\sigma} \partial^\nu
b^{\rho\sigma},
\eeq
which is invariant under the gauge transformation $b_{\mu\nu}
\longrightarrow b_{\mu\nu} +\partial_\mu b_\nu -\partial_\nu
b_\mu$. The expansion (\ref{linexp}) shows that the linear
multiplet contains a real scalar field $C$, a Majorana spinor
$\chi$ and an antisymmetric tensor $b_{\mu\nu}$, with the same
physical dimension as $C$. In contrast with vector or chiral
superfields, the linear multiplet {\it does not} possess auxiliary
fields, a fact which will have immediate consequences for
supersymmetry breaking.

Since the linear multiplet is a constrained vector superfield,
a supersymmetric action involving $L$ and chiral superfields
$\Sigma^i$ is constructed with an integral over the full
superspace:
\beq
\label{action1}
S = \int d^4x\, {\cal L}= 2\int d^4x \int d^2\theta d^2\ov
\theta\, \Phi(L,\Sigma^i, \ov\Sigma^i),
\eeq
where $\Phi$ is a real function (a real vector superfield).
It is well known that a supersymmetric duality transformation
[\ref{S}] can always be performed to transform $L$ into a
chiral multiplet so that theory (\ref{action1}) is (classically)
equivalent to a particular supersymmetric non-linear $\sigma$-model
with chiral multiplets only.

The linear multiplet becomes more interesting when the general
coupling of linear and chiral multiplets is submitted to gauge
invariance. We will for simplicity consider a unique linear
multiplet $L$, chiral multiplets $\Sigma^i$ transforming in
some representation of the gauge group, and
gauge fields described by a vector superfield $V= V^AT^A$,
$T^A$ being generators of the gauge group in the representation
of the chiral multiplets, with normalization
$$
\Tr (T^AT^B) = \tau \delta^{AB}.
$$
With chiral multiplets only, one would write a general
supersymmetric gauge invariant lagrangian
\beq
\label{chiralS}
\begin{array}{rcl}
{\cal L}_\Sigma &=&
\bigint d^2\theta d^2\ov\theta\,  K(\Sigma^i , (\ov\Sigma e^V)_i) \crbig
&&
+\bigint d^2\theta\, F(\Sigma^i, W^{\alpha A})
+\bigint d^2\ov\theta\, \ov F(\ov\Sigma_i, \ov W^{\dalpha A})
\end{array}
\eeq
where $W^{\alpha A}$ are the field strength superfields,
\beq
\label{Walpha}
W^\alpha = -{1\over4}\ov{\cal DD}e^{-V}{\cal D}^\alpha e^V =
W^{\alpha A}T^A.
\eeq
There are two arbitrary functions, $K$ which is real and $F$
which is a function of chiral superfields only, both must be
gauge invariant. Chiral kinetic terms will be generated by
$K$ while $F$ describes in particular gauge kinetic terms.
One easily verifies that lagrangian (\ref{chiralS}) does not
contain terms with more than two derivatives. The minimal
choice of gauge kinetic lagrangian corresponds to the most
common choice
\beq
\label{usualF}
F(\Sigma^i, W^\alpha) = w(\Sigma^i) + {1\over4}f_{AB}(\Sigma^i)
W^{\alpha A} {W_\alpha}^B,
\eeq
where $w(\Sigma^i)$ is the superpotential. This minimal form
leads to the following gauge kinetic terms:
\beq
\label{gaugekin1}
-{1\over4} [f_{AB}(z^i) + \ov f_{AB}(\ov z_i)] F^{A\,\mu\nu}
F^B_{\mu\nu},
\eeq
the complex scalar fields $z^i$ being the lowest components of
superfields $\Sigma^i$.
Since the linear multiplet is not chiral, it cannot appear in the
chiral density which defines gauge kinetic terms. On the contrary,
the function $K$ can freely depend on $L$ since $K$ is a real
vector superfield. Adding a gauge invariant linear multiplet
$L$ to theory (\ref{action1}) could then proceed by the
replacement
$$
K(\Sigma^i , (\ov\Sigma e^V)_i) \longrightarrow K(L, \Sigma^i ,
(\ov\Sigma e^V)_i).
$$
The resulting lagrangian would always include gauge kinetic
terms with a harmonic metric, as in eq. (\ref{gaugekin1}).

The interesting point is that this simple solution is not unique.
There exists the possibility of a coupling of a gauge variant
linear multiplet to chiral multiplets and gauge fields which
also escapes the condition of gauge kinetic terms with harmonic
metric. This situation cannot be
achieved with lagrangian (\ref{chiralS}) as starting point.
We firstly need some manipulations of (\ref{chiralS}) with the
minimal choice (\ref{usualF}) to obtain a form more appropriate
to the introduction of the linear multiplet.
Suppose we define the real vector superfield $\Omega(V)$ by the
conditions
\beq
\label{CSdef}
\ov{\cal DD} \Omega(V) = \tau^{-1}\Tr(W^\alpha W_\alpha),  \qquad
{\cal DD} \Omega(V) = \tau^{-1}\Tr(\ov W_\dalpha \ov W^\dalpha),
\eeq
then, for an arbitrary chiral function $f$,
\beq
\label{manip}
\begin{array}{rcl}
{1\over4\tau}{\displaystyle\int} d^2\theta\, f \Tr W^\alpha W_\alpha
+ {\rm h.c.} &=& {1\over4}{\displaystyle\int} d^2\theta\,
\ov{\cal{DD}} [f\Omega(V)]+ {\rm h.c.}\crbig
&=& - {\displaystyle\int} d^2\theta d^2\ov\theta\,
(f+\ov f)\Omega(V) \crbig
&=&
{1\over2} {\displaystyle\int} d^2\theta d^2\ov\theta\,
(f+\ov f)[L-2\Omega(V)].
\end{array}
\eeq
The last step uses $\int d^4x\int d^2\theta d^2\ov\theta\,
(f+\ov f)L=0$ since $f$ is chiral and $L$ linear.
These manipulations allow to move gauge kinetic terms from a
chiral lagrangian into a $D$-density provided one replaces the
chiral gauge invariant superfield $\tau^{-1}\Tr(WW)$ by the Chern-Simons
superfield $\Omega(V)$.

The Chern-Simons superfield $\Omega(V)$, defined by conditions
(\ref{CSdef}), is not gauge invariant. Its definition indicates
however that its gauge transformation $\delta\Omega$ is a
linear superfield:
\beq
{\cal DD} \delta\Omega = \ov{\cal DD} \delta\Omega = 0.
\eeq
Gauge invariance of the last expression (\ref{manip}) follows
from this observation.
We now impose that the gauge transformation of the linear
superfield is
\beq
\label{Ltransf}
\delta L = 2 \delta\Omega(V).
\eeq
Since the combination
\beq
\label{Lhat}
\hat L = L - 2 \Omega(V)
\eeq
is by construction gauge invariant, the supersymmetric lagrangian
\beq
\label{L1}
{\cal L}_L = 2\int d^2\theta d^2\ov\theta\, \Phi (\hat L, \Sigma^i,
(\ov\Sigma e^V)_i)
+\int d^2\theta\, w(\Sigma^i) + \int d^2\ov\theta\, \ov
w(\ov\Sigma_i)
\eeq
is also gauge invariant for any real function $\Phi$. According
to equalities (\ref{manip}), it is actually a generalization of
lagrangian (\ref{chiralS}) with the usual gauge sector
(\ref{usualF}). The component expansion of (\ref{L1}) contains
gauge kinetic terms of the form
$$
-{1\over2} \left[{\partial\over\partial L}\Phi(L, \Sigma^i,
\ov\Sigma_i) \right]_{\theta=\ov\theta=0}F^A_{\mu\nu} F^{A\,\mu\nu},
$$
with a non-harmonic metric in general. It has been realized
[\ref{DFKZ1}] that this form of linear multiplet coupling is
important to describe quantum corrections to orbifold effective
supergravities.

The Chern-Simons superfield can be explicitly computed by solving
eqs. (\ref{CSdef}) with the help of (\ref{Walpha}) and Bianchi
identities. In the non-abelian case, its expression is complicated
(see for instance ref. [\ref{CFV}]). The simpler abelian
Chern-Simons superfield reads
\beq
\label{CSabel}
\Omega(V) = -{1\over4} \tau^{-1} \Tr
\left[ ({\cal D}^\alpha V)W_\alpha +
(\ov{\cal D}_\dalpha V)\ov W^\dalpha + V({\cal D}^\alpha
W_\alpha)\right].
\eeq
Its component expansion is given in the appendix.

The component expansion of lagrangian (\ref{L1}) is relatively
cumbersome to obtain. Since one of our goals is to discuss the
sector of the theory which controls supersymmetry breaking with
and without gaugino condensation, we will only need to obtain
the scalar and gaugino lagrangian. For simplicity, we will consider
below the lagrangian (\ref{L1}) for a unique, gauge-invariant
chiral multiplet $\Sigma$, with components $(z,\psi,f)$. We will
truncate the chiral superfield $\Sigma$ and omit the fermion
$\psi$ which is irrelevant to our discussion. But the auxiliary
field $f$ must be retained. Analogously, the linear superfield
$L$, with component expansion (\ref{linexp}) will be truncated
by keeping only the real scalar $C$ and the transverse vector
$v_\mu$, equivalent to the antisymmetric tensor $b_{\mu\nu}$
and by duality to a pseudoscalar. Finally, all gauge boson
contributions will be omitted. In this situation and using
the Wess-Zumino gauge, all contributions of gauginos $\lambda^A$
and auxiliary
fields $D^A$ are obtained using the abelian Chern-Simons superfield
(\ref{CSabel}), even for a non-abelian gauge group. We can
directly assume that gauge auxiliary fields $D^A$ vanish.
Their couplings to the linear multiplet only proceed through
terms of the form $D^A(\lambda^A\chi)$ which are omitted
here. With non-singlet chiral matter (and with $\psi=0$),
the auxiliary fields $D^A$ would only induce the usual
positive potential term which is not of central importance
when discussing supersymmetry breaking.
In lagrangian (\ref{L1}), the real function $\Phi$ is
arbitrary up to constraints related to the positivity of
kinetic energy, which will be stated using  the component
expansion of the lagrangian. The truncated lagrangian
obtained by selecting only scalar, $v_\mu$, $f$ and gaugino
contributions is
\beq
\label{lagrancomp}
\begin{array}{rcl}
{\cal L}_L &=& {1\over2}\Phi_{CC}(\partial_\mu C)(\partial^\mu C)
- {1\over2}\Phi_{CC} v_\mu v^\mu
- 2\Phi_{\Sigma\ov \Sigma} (\partial_\mu z)(\partial^\mu \ov z) \crbig
&&+  v_\mu\left[-i\Phi_{C\Sigma}(\partial^\mu z) +
i\Phi_{C\ov \Sigma}(\partial^\mu \ov z)
+ \Phi_{CC}(\lambda^A\sigma^\mu\ov\lambda^A) \right]
\crbig
&& - \Phi_C\left[i\lambda^A\sigma^\mu\partial_\mu\ov
\lambda^A
-i \partial_\mu\lambda^A\sigma^\mu\ov \lambda^A \right]
\crbig
&& +i (\lambda^A\sigma^\mu\ov\lambda^A)
\left[ \Phi_{C\Sigma}(\partial_\mu z)
- \Phi_{C\ov \Sigma}(\partial_\mu
\ov z) \right] \crbig
&& +{1\over2}\Phi_{CC}\left[ (\lambda^A\lambda^A)(\ov\lambda^B
\ov\lambda^B)
+2(\lambda^A\lambda^B)(\ov\lambda^A\ov\lambda^B)\right]
\crbig
&& + {\cal L}_{AUX},
\end{array}
\eeq
where we use the notation
$$
\begin{array}{c}
\Phi_C = \left[{\partial\Phi\over\partial L}\right]_
{\theta=\ov\theta=0}
={\partial\over\partial C}
\Phi(C,z,\ov z), \crbig
\Phi_{C\Sigma} = \left[{\partial^2\Phi\over\partial
L\partial \Sigma}\right]_{\theta=\ov\theta=0}
= {\partial^2\over\partial
C\partial z}\Phi(C,z,\ov z),
\end{array} \quad\ldots,
$$
and the auxiliary field lagrangian is
\beq
\label{Laux}
{\cal L}_{AUX} = 2\Phi_{\Sigma\ov \Sigma} f\ov f
- f\Phi_{C\Sigma}(\lambda^A\lambda^A) -\ov f
\Phi_{C\ov \Sigma}(\ov\lambda^A\ov\lambda^A) + f{dw\over dz}
+\ov f {d\ov w\over d\ov z}.
\eeq
Notice that scalar kinetic terms in (\ref{lagrancomp}) do not
mix $z$, $C$ and $v_\mu$. Positivity of the kinetic terms for
$\lambda^A$, $C$ (and $v_\mu$) and $z$ respectively corresponds
to the conditions
\beq
\label{positive1}
\Phi_C > 0, \qquad
\Phi_{CC}< 0, \qquad \Phi_{\Sigma\ov \Sigma} > 0
\eeq
in the domains of the scalar fields. The equation of motion for
$f$ is
\beq
\label{feom}
f = -{1\over2}{\Phi_{\Sigma\ov \Sigma}}^{-1}\left[ {d\ov w\over d\ov z}
- \Phi_{C\ov \Sigma} (\ov\lambda^A\ov\lambda^A)\right].
\eeq
In the absence of gaugino condensates, $\langle\lambda^A\lambda^A\rangle
= 0$, the scalar potential is
\beq
\label{pot1}
V = {1\over2}{\Phi_{\Sigma\ov \Sigma}}^{-1} \left| {dw\over dz} \right|^2,
\eeq
which is semi-positive definite according to conditions (\ref{positive1}).
If
there is a vacuum $\langle z\rangle$ such that
$\langle{dw\over dz}\rangle=0$, $\vev{f}=\vev{V}=0$
and supersymmetry is not broken. The linear multiplet does not
play any
r\^ole since the superpotential is independent of $L$ and $L$
does not possess any auxiliary field.

\subsection{Duality transformation}

The antisymmetric tensor contained in the linear multiplet $L$
can always be transformed into a pseudoscalar field. In the
supersymmetric context, this duality transformation can be
performed at the superfield level, the linear multiplet being
replaced by a chiral multiplet $S$. To perform the supersymmetric
duality transformation [\ref{S}], replace the lagrangian (\ref{L1}) by
\beq
\label{L2}
\begin{array}{rcl}
{\cal L}_U &=&
2\bigint d^2\theta d^2\ov\theta\, \left[ \Phi(U,\Sigma,\ov\Sigma) -
(S+\ov S)(U+2\Omega(V)) \right]
+\bigint d^2\theta\,  w(\Sigma) + \bigint d^2\ov\theta\, \ov
w(\ov\Sigma) \crbig
&=& 2\bigint d^2\theta d^2\ov\theta\, \left[ \Phi(U,\Sigma,
\ov\Sigma) - (S+\ov S)U\right]
\crbig
&& +\bigint d^2\theta\, \left[  SW^AW^A + w(\Sigma)
\right]
+\bigint d^2\ov\theta\, \left[ \ov S\ov W^A\ov W^A +
\ov w(\ov\Sigma) \right],
\end{array}
\eeq
where $U$ is an unconstrained vector superfield. Notice that
the $d^2\theta d^2\ov\theta$ integral does not depend on the
gauge superfield $V$. The equation of motion for the chiral
multiplet $S$ implies that $U+2\Omega(V)$ is a
linear multiplet and ${\cal L}_U$ is then equivalent with
${\cal L}_L$. On the other hand, the equation of motion for
$U$ is
\beq
\label{Ueom}
{\partial \Phi\over\partial U} = S+\ov S,
\eeq
which can in principle be inverted to express the vector
superfield $U$ as a function of $S+\ov S$, $\Sigma$ and
$\ov\Sigma$. Inserting the expression $U(S+\ov S, \Sigma,
\ov\Sigma)$ into lagrangian (\ref{L2}) leads to
\beq
\label{L3}
\begin{array}{rcl}
{\cal L}_S &=& \bigint d^2\theta d^2\ov\theta\, K(S+\ov S,
\Sigma, \ov\Sigma) \crbig
&& +\bigint d^2\theta\, \left[ SW^AW^A + w(\Sigma)
\right]
+\bigint d^2\ov\theta\, \left[ \ov S\ov{W^AW^A} +
\ov w(\ov\Sigma)\right] ,
\end{array}
\eeq
where
\beq
\label{Kdef}
K(S+\ov S, \Sigma, \ov\Sigma) = 2\left[\Phi-(S+\ov S)U
\right]_{ U= U(S+\ov S,\Sigma,\ov\Sigma)}.
\eeq
In components, the supersymmetric duality transformation works
in the following way. Denoting the components of the
unconstrained vector superfield $U$ by
$$
U \, : \, (C,\varphi, m,n, \omega_\mu, \eta, d),
$$
and the components of the chiral multiplet $S$ by
$$
S \, : \, (s, \psi_s, f_s),
$$
the highest component $d$ appears in ${\cal L}_U$ in
$$
{\partial\Phi\over\partial C}d - (s+\ov s)d.
$$
The equation of motion for $d$ defines then the real part of
$S$ as
\beq
\label{reals}
s + \ov s = \Phi_C,
\eeq
which is the lowest component of the superfield equation
(\ref{Ueom}). By inverting this relation, one can express
$C$ as a function of $s+\ov s$ and $z$, the scalar component
of the chiral multiplet $\Sigma$.

The spinor $\eta$ which appears in the $\ov{\theta\theta}\theta$
component of $U$ also contributes to ${\cal L}_U$ like a Lagrange
multiplier. Its equation of motion, obtained from the $\theta$
component of eq. (\ref{Ueom}), gives the definition of the
fermionic component $\psi_s$ of $S$ as a function of the
components of $U$ and $\Sigma$:
\beq
\label{psis}
\psi_s = {i\over\sqrt2} \Phi_{CC} \varphi + \Phi_{C\Sigma}
\psi .
\eeq
This equation is actually used to eliminate $\varphi$ which
can be expressed as a function of $\psi_s$, $\psi$, $s+\ov s$
and $z$, using also (\ref{reals}).

The scalar fields $m$ and $n$, which correspond to the
$\theta\theta$ and $\ov{\theta\theta}$ components of $U$,
appear quadratically and without derivatives in ${\cal L}_U$.
Their equations of motion are, in complex form,
\beq
\label{mandn}
f_s = {i\over2} \Phi_{CC} (m+in) + \Phi_{C\Sigma} f
+ {1\over4}\Phi_{CCC}(\varphi\varphi)
-{i\over\sqrt2}\Phi_{CC\Sigma}(\varphi\psi) - {1\over2}
\Phi_{C\Sigma\ov\Sigma} (\psi\psi),
\eeq
which define $f_s$, the auxiliary component of $S$. It can
be used to eliminate $m+in$. Eq. (\ref{mandn}) is the
$\theta\theta$ component of the superfield equation
(\ref{Ueom}).

The vector component of $U$, $\omega_\mu$, which also has
an algebraic equation of motion in ${\cal L}_U$, leads to
the usual duality transformation. Its equation of motion is
\beq
\label{omega}
\begin{array}{rcl}
\partial_\mu{\Im s} &=& {1\over2}\Phi_{CC}\omega_\mu
- {i\over2}\Phi_{C\Sigma}\partial_\mu z
+{i\over2} \Phi_{C\ov \Sigma}\partial_\mu \ov z
+{1\over4}\Phi_{CCC}(\varphi\sigma_\mu\ov\varphi) \crbig
&& - {i\over2\sqrt2} \Phi_{CC\Sigma} (\psi\sigma_\mu\ov\varphi)
+{i\over2\sqrt2} \Phi_{CC\ov\Sigma}(\varphi\sigma_\mu\ov\psi)
+{1\over2} \Phi_{C\Sigma\ov\Sigma} (\psi\sigma^\mu\ov\psi).
\end{array}
\eeq
The component $\omega_\mu$ can then be replaced by the
derivative of the imaginary part of $s$.

Finally, components $C$ and $\varphi$ are propagating
fields in ${\cal L}_U$. But they can be replaced by $\Re s$
and $\psi_s$ using eqs. (\ref{reals}) and (\ref{psis}).

In terms of the components of chiral multiplets $S$ and
$\Sigma$, the truncated lagrangian can be written
\beq
\label{LScomp}
\begin{array}{rcl}
{\cal L}_S &=&
-2{\Phi_{CC}}^{-1}\left[ -(\partial_\mu s)(\partial^\mu \ov s)
+ f_s\ov f_s\right]
\crbig
&& +2\left[ \Phi_{\Sigma\ov\Sigma} - {\Phi_{CC}}^{-1}\Phi_{C\Sigma}
\Phi_{C\ov\Sigma}\right] \left[ -(\partial_\mu z)
(\partial^\mu \ov z) + f\ov f\right] \crbig
&& +2{\Phi_{CC}}^{-1}\Phi_{C\Sigma}\left[ -(\partial_\mu\ov s)
(\partial^\mu z) + \ov f_s f\right] \crbig
&& +2{\Phi_{CC}}^{-1}\Phi_{C\ov\Sigma}\left[ -(\partial_\mu s)
(\partial^\mu\ov z) + f_s \ov f\right] \crbig
&& -2i s\lambda^A\sigma^\mu\partial_\mu\ov\lambda^A
+2i \ov s\partial_\mu \lambda^a\sigma^\mu\ov\lambda^A
\crbig
&& -f_s(\lambda^A\lambda^A) -\ov
f_s(\ov\lambda^A\ov\lambda^A) +{dw\over dz}f +
{d\ov w\over d\ov z}\ov f.
\end{array}
\eeq
In this expression, $\Phi$ and its derivatives should be considered
as functions of $s+\ov s$, $z$ and $\ov z$, using relation
(\ref{reals}) to eliminate $C$.
The metric of scalar kinetic terms is not diagonal. Its
determinant is $-4{\Phi_{CC}}^{-1} \Phi_{\Sigma\ov\Sigma}$
and positivity of the kinetic terms leads again to
conditions (\ref{positive1}). In particular, positivity
of gauge kinetic terms corresponds to
\beq
\label{positive2}
\Re s > 0,
\eeq
in view of eqs. (\ref{reals}) and (\ref{positive1}).
Notice that the auxiliary field contributions can be
rearranged into
\beq
\label{Lsaux}
\begin{array}{rcl}
{\cal L}_{AUX} &=&
-2{\Phi_{CC}}^{-1} \tilde f\ov{\tilde f} + 2\Phi_{\Sigma\ov
\Sigma}f\ov f
+\tilde f(\lambda^A\lambda^A)
+\ov{\tilde f} (\ov\lambda^A\ov\lambda^A) \crbig
&& + f\left[ {dw\over dz} -\Phi_{C\Sigma}(\lambda^A
\lambda^A)\right]
+ \ov f\left[ {d\ov w\over d\ov z}-\Phi_{C\ov\Sigma}
(\ov\lambda^A\ov\lambda^A)\right],
\end{array}
\eeq
where $\tilde f = -f_s+\Phi_{C\Sigma}f$. This redefinition
diagonalises the auxiliary field equations of motion which
read
\beq
\label{auxeom}
\begin{array}{rcl}
\tilde f &=& {1\over2} \Phi_{CC} (\ov\lambda^A\ov\lambda^A),
\crbig
f &=& -{1\over2}{\Phi_{\Sigma\ov\Sigma}}^{-1} \left[{d\ov w\over d\ov z} -
\Phi_{C\ov\Sigma}(\ov\lambda^A\ov\lambda^A)\right].
\end{array}
\eeq
The auxiliary field $f_s$ is
\beq
\label{fseom}
\begin{array}{rcl}
f_s &=& \Phi_{C\Sigma}f - {1\over2}\Phi_{CC}(\ov\lambda^A\ov
\lambda^A) \crbig
&=& -{1\over2}{\Phi_{\Sigma\ov\Sigma}}^{-1}\Phi_{C\Sigma}{d\ov w\over
d\ov z}
-{1\over2}\left[ \Phi_{CC} -
{\Phi_{\Sigma\ov\Sigma}}^{-1}\Phi_{C\Sigma}\Phi_{C\ov\Sigma}
\right] (\ov\lambda^A\ov\lambda^A).
\end{array}
\eeq
The scalar potential takes the simple form
\beq
\label{pots}
V = - 2{\Phi_{CC}}^{-1}\tilde f\ov{\tilde f} + 2\Phi_{\Sigma\ov
\Sigma}f\ov f,
\eeq
with $\tilde f$ and $f$ replaced by the scalar contributions
in the expressions (\ref{auxeom}). It is positive or zero as
a consequence of the positivity of kinetic terms.

If gauginos do not condense, $\langle\lambda^A\lambda^A\rangle = 0$, the
scalar parts of the auxiliary fields reduce to
\beq
\begin{array}{rcl}
f &=& -{1\over2}{\Phi_{\Sigma\ov\Sigma}}^{-1} {d\ov w\over d\ov z},
\crbig
f_s &=& \Phi_{C\Sigma}f,
\end{array}
\eeq
and $\tilde f= 0$. The new auxiliary field $f_s$, defined by
the supersymmetric duality transformation of the linear
multiplet is simply proportional to $f$ \footnote
{Except if $\Phi_
{C\Sigma}=0$, which implies
$\Phi = F(\hat L) + G(\Sigma, \ov\Sigma).$}.
Clearly, supersymmetry is unbroken whenever
$\langle{dw\over dz}\rangle=0$.
Without gaugino condensation, the scalar potential
$$
V = {1\over2}{\Phi_{\Sigma\ov\Sigma}}^{-1} \left|{dw\over dz}\right|^2
$$
is of course the same as in the theory expressed with the linear
multiplet, in eq. (\ref{pot1}). Its dependence on $s+\ov s$ is
entirely in the factor ${\Phi_{\Sigma\ov\Sigma}}^{-1}$. Then, if
the $s$-independent condition ${dw\over dz}=0$ has a solution,
the scalar potential has a flat direction leaving $s+\ov s$
undetermined.

\subsection{Gaugino condensation}

To discuss the effect of gaugino condensation, we use an
expectation value
\beq
\langle\lambda^A\lambda^A \rangle = \Lambda^3
\eeq
inserted in the three equivalent lagrangians ${\cal L}_L$
[eq. (\ref{L1})], ${\cal L}_U$ [eq. (\ref{L2})] and ${\cal L}_S$
[eq. (\ref{L3})]. In ${\cal L}_U$ and ${\cal L}_S$, the gauge
multiplet only appears in $\int d^2\theta\, SW^AW^A +
{\rm h.c}$. Since
\beq
\begin{array}{rcl}
W^AW^A &=& -\lambda^A\lambda^A - 2i\theta\lambda^AD^A +
(\theta\sigma^\mu\ov\sigma^\nu\lambda^A) F_{\mu\nu}^A \crbig
&& + \theta\theta\left[
-2i\lambda^A\sigma^\mu\partial_\mu\ov\lambda^A + D^AD^A -
{1\over2}F^A_{\mu\nu} F^{A\,\mu\nu}
- {i\over4}\epsilon^{\mu\nu\rho\sigma}F^A_{\mu\nu}
F^A_{\rho\sigma}
\right],
\end{array}
\eeq
the insertion of a non-zero $\langle\lambda^A\lambda^A\rangle$ can be
regarded as a shift of the superfield
$$
W^AW^A \longrightarrow W^AW^A - \Lambda^3.
$$
Equivalently, the theory with gaugino condensation has a
superpotential
\beq
\label{neww}
w_\Lambda(S, \Sigma) = w(\Sigma) - \Lambda^3 S.
\eeq
Since eq. (\ref{neww}) is a superfield equation, the resulting
lagrangian, with modified superpotential $w_\Lambda$, is
supersymmetric. Supersymmetry breaking only occurs if the
theory does not possess a supersymmetry invariant vacuum,
i.e. supersymmetry breaking would seem spontaneous.

The appearance of a contribution $-\Lambda^3 S$ in
the superpotential corresponds to the addition of
$$
-\Lambda^3 (f_s+\ov f_s)
$$
to the component lagrangian (\ref{LScomp}). As it should, this
is the same as replacing $\lambda^A\lambda^A$ by $\lambda^A\lambda^A
+ \Lambda^3$ directly in eq. (\ref{LScomp}).
The superpotential (\ref{neww}) does not generate any mass
term for the fermionic component $\psi_s$ of $S$, which is
massless for arbitrary expectation values of the scalar fields.
The expectation
values of auxiliary fields become
\beq
\label{auxLambda}
\begin{array}{rcl}
\vev f &=& -\langle {1\over2}{\Phi_{\Sigma\ov\Sigma}}
^{-1}\left[ {d\ov w\over d\ov z}
- \Phi_{C\ov\Sigma}\Lambda^3 \right] \rangle, \crbig
\langle f_s\rangle &=& \langle \Phi_{C\Sigma}f -
{1\over2}\Phi_{CC}\Lambda^3 \rangle,
\end{array}
\eeq
which shows that supersymmetry is always broken since $\vev f$
and $\vev{f_s}$ may simultaneously vanish only if $\Lambda=0\,$\footnote
{The case $\langle\Phi_{CC}\rangle =0$ is singular, according to
eq. (\ref{LScomp}), which means that the duality transformation
cannot be performed.}.
The potential with $\Lambda \ne 0$ is
\beq
\label{potLambda}
V_\Lambda = {1\over2}{\Phi_{\Sigma\ov\Sigma}}^{-1} \left|{dw\over dz}
- \Lambda^3\Phi_{C\Sigma} \right|^2 - {1\over2}\Phi_{CC}
\Lambda^6.
\eeq
With positivity conditions (\ref{positive1}), the scalar
potential is strictly positive, another indication that
supersymmetry is broken. It is plausible that $V_\Lambda$
will find its minimum at $\vev f = 0$, a
situation characterized by
\beq
\label{Lambdavacuum}
\begin{array}{rcl}
\vev f &=& 0  \qquad\Longrightarrow\qquad \langle{dw\over dz}\rangle
= \langle\Phi_{C\Sigma}\rangle \Lambda^3,
\crbig
\vev{f_s} &=& -{1\over2} \langle\Phi_{CC}\rangle \Lambda^3 \ne 0.
\end{array}
\eeq
If possible, the superpotential will adjust itself in
order that supersymmetry breaking is entirely in the
$S$-sector. Notice also that supersymmetry breaking will
in general destroy the flat direction and lift the
degeneracy in $s+\ov s$.

In the version of the theory using the linear multiplet,
with lagrangian (\ref{L1}), the components of the gauge
multiplet appear in the Chern-Simons superfield associated
with $L$ in the combination $\hat L=L-2\Omega(V)$.
The superfield $\Omega(V)$ contains gaugino bilinear
$\lambda^A\lambda^A$ in its $\theta\theta$ and $\ov\theta\ov\theta$
components only. This is consistent with the definition of
$\Omega(V)$, through the condition $\ov{\cal DD}\Omega(V) =
\tau^{-1}W^AW^A$
and its conjugate. The insertion of the expectation value of
gaugino bilinears would correspond to replacing $\Omega(V)$ by
\beq
\label{OmegaLambda}
\Omega_\Lambda = \Omega(V) +{1\over4}\Lambda^3 \left[ \theta\theta
+ \ov\theta\ov\theta \right],
\eeq
which is clearly not a superfield equation. This replacement
would then be interpreted as an explicit breaking of supersymmetry.
In the component expansion of the lagrangian, eq. (\ref{lagrancomp}),
the insertion of the gaugino condensate leads to the lagrangian
\beq
{\cal L}_\Lambda = {\cal L}_L - \Phi_{C\Sigma}f\Lambda^3
- \Phi_{C\ov\Sigma}\ov f\Lambda^3 + {1\over2} \Phi_{CC}
\Lambda^6.
\eeq
The expectation value of the unique auxiliary field $f$ becomes
\beq
\label{feom2}
\vev f = -\langle{1\over2}{\Phi_{\Sigma\ov\Sigma}}^{-1} \left[
{d\ov w\over d\ov z} -
 \Phi_{C\ov\Sigma} \Lambda^3 \right]\rangle,
\eeq
as in the first eq. (\ref{auxLambda}). And the scalar potential
is identical to eq. (\ref{potLambda}) which is strictly
positive \footnote{
The case $\langle \Phi_{CC}\rangle =0$ corresponds, according to eq.
(\ref{lagrancomp}), to a non-propagating linear multiplet.
}.
In the formulation with the linear multiplet, the insertion of
gaugino condensates is an explicit breaking of supersymmetry,
generating in particular a positive contribution to the scalar
potential. This mechanism does not use auxiliary fields which
are absent in the linear multiplet.

\nsect{Supergravity with a linear multiplet: \hfill\break preliminaries}

The case of global supersymmetry with a linear multiplet presented in
the previous section, even if suggestive, is not sufficient to discuss
in general terms the low-energy effective supergravity theories
obtained from superstrings. We have to consider the general
coupling of a linear multiplet   to supergravity. This problem
has been studied, to a large extent, in the literature using either
superconformal methods [\ref{FGKVP}, \ref{CFV}] or super-Poincar\'e
superspace techniques [\ref{Oetal}, \ref{Betal}, \ref{BGG}, \ref{ABGG}]
but the general expression for the lagrangian in terms of arbitrary
functions is not available, contrary to the case of
chiral fields coupled to supergravity [\ref{CFGVP}]. In this section and
the next,  we present such general expressions  not for the
full lagrangian but for the terms that interest us  the most, {\it i.e.}
the scalar and gauge kinetic terms as well as the scalar
potential including gaugino bilinears. We will explain the limitations
that make this calculation more difficult than the one
in ref. [\ref{CFGVP}] and obtain the latter as a particular case.

In the following, we will use the formalism of superconformal
supergravity [\ref{KTPVN}] which appears
to be the most appropriate for our
purposes. This method has the advantage that the supergravity
lagrangian (Einstein and Rarita-Schwinger terms) is always
included in matter couplings, through
covariantization of derivatives and invariant density formul\ae.
Moreover it nicely keeps track of the breaking of scale symmetry
which will be an important issue when discussing the
renormalization-group behaviour in the effective low-energy
theory. The formalism of superconformal supergravity is
reviewed in refs. [\ref{KU}], and we will use the same
conventions\footnote{As in section 2,
the space-time metric has signature $(-,+,+,+)$
in our supergravity expressions.}.
The general idea is to firstly construct an
action invariant under the transformations of the superconformal
algebra. This theory describes all matter and vector multiplets,
but it also depends on an additional multiplet called compensator.
As a gauge theory of the superconformal algebra, it includes
the gauge potentials $(e^m_\mu, \omega_\mu^{mn}, f_\mu^m,
b_\mu, \psi_\mu, \varphi_\mu, A_\mu)$. The gauge fields
$\omega_\mu^{mn}$ (Lorentz transformations), $\varphi_\mu$
(special supersymmetry) and $f_\mu^m$ (conformal boosts)
can be algebraically solved as a consequence of the imposition
of constraints on the curvatures. The second step is to obtain
a super-Poincar\'e theory by choosing a gauge for conformal
boosts, dilatations, chiral $U(1)$ and special supersymmetry.
This is done by assigning specific field-dependent values to
certain components of the compensator and to some gauge fields
of the superconformal algebra, the exact procedure depending
on the choice of compensating multiplet.

The choice of the compensator dictates the set of auxiliary
fields present in the supergravity multiplet of the Poincar\'e
theory, which is not unique. It is known [\ref{FGKVP}] that the
simplest choice of a chiral compensating multiplet with Weyl
and chiral weights equal to one, which leads to `minimal
supergravity', also leads to the most general class of
matter--supergravity couplings. Denoting the components of
the chiral compensator $S_0$ by
\beq
S_0 : (z_0, \psi_0 , f_0),
\eeq
the standard procedure [\ref{KU2}] for fixing the superconformal
symmetries is to choose a gauge in which conformal boosts are
fixed by imposing that the gauge field of dilatations $b_\mu$
vanishes, special supersymmetry and chiral $U(1)$ are fixed by
choosing a specific field-dependent form of the component
$\psi_0$ of $S_0$ and by the condition $\Im z_0=0$. Finally,
the gauge fixing of dilatations corresponds to choosing $|z_0|$
in such a way that the Einstein term in the theory has the
canonical form
$$
-{1\over2}{1\over\kappa^2} eR,
$$
where $e$ is the vierbein determinant and $R$ the curvature
scalar.
This gauge fixing procedure applied to the chiral compensator
$S_0$ leaves only $f_0$ (a complex scalar) and $A_\mu$ (the
gauge potential of chiral $U(1)$ transformations) unspecified.
They will be the auxiliary fields of minimal Poincar\'e
supergravity.

\subsection{The chiral case}

To illustrate the construction, we first consider the
most general lagrangian density for one chiral multiplet
$\Sigma$, with a gauge multiplet $V=V^AT^A$. Both multiplets can
be taken with zero Weyl and chiral weights without loss of
generality. This is a particular case of the theory
constructed in ref. [\ref{CFGVP}]. Using supermultiplet
expressions, the lagrangian is
\beq
\label{Cremmer1}
{\cal L}= -{3\over2}\big[S_0\ov S_0 e^{-{1\over3}K(\Sigma,
\ov\Sigma e^V)} \big]_D
+\big[{1\over4} f(\Sigma) W^AW^A  + S_0^3 w(\Sigma)\big]_F
\eeq
where $S_0$ is the chiral compensator, $K$ and $f$ are
arbitrary functions ($K$ is real and $f$ analytic), $W^A$
is the chiral supermultiplet obtained from $V^A$ which contains
gauge curvatures (the local analog of the superfield $W^{A\alpha}$
of global supersymmetry) and $w(\Sigma)$ is the superpotential.
This expression only makes sense in the context of tensor
calculus, which specifies the rules for combining supergravity
multiplets. The two terms correspond to
D- and F-density formulae for invariant actions. A D-density
is an invariant action obtained by combining the components
of a real vector multiplet $\cal V$ with conformal weight
two and superconformal gauge fields. In eq. (\ref{Cremmer1}),
${\cal V} = -{3\over2} S_0\ov S_0 e^{-{1\over3}K(\Sigma,
\ov\Sigma e^V)}$. The condition on the Weyl weight and
reality determine the form of compensator contributions.
In the same way, an F-density combines components of a
chiral multiplet $\cal S$ with Weyl and chiral weights
equal to three with gauge fields to form an invariant
action. Again, ${\cal S}=  {1\over4} f(\Sigma) W^AW^A  +
S_0^3 w(\Sigma)$ has by construction the correct
weights.

Omitting all gravitino contributions in the density formula,
one has
\beq
\label{densities}
\begin{array}{rcl}
[{\cal V}]_D &=& e(d+{1\over3}cR), \\
{[{\cal S}]_F} &=& ef + {\rm h.c.},
\end{array}
\eeq
where ${\cal V}$ is a vector multiplet with components $(c,
\chi, m, n, b_\mu, \lambda, d)$ and ${\cal S}$ a chiral
multiplet with components $(z,\psi,f)$.
The second term in the D-density
formula contains the curvature scalar. The components $d$,
$c$ and $f$ of the multiplets
\beq
\begin{array}{rcl}
{\cal V} &=&  -{3\over2} S_0\ov S_0 e^{-{1\over3}K(\Sigma,
\ov\Sigma e^V)} , \crbig
{\cal S} &=& {1\over4} f(\Sigma) W^AW^A + S_0^3 w(\Sigma),
\end{array}
\eeq
can be calculated using the rules of superconformal tensor
calculus [\ref{KU}].
It turns out that neither $d$ nor $f$ contain contributions
involving the curvature scalar $R$. Then, the Einstein
lagrangian appears in
\beq
\label{ECremmer}
{1\over3} ceR = -{1\over2}eR[ z_0\ov z_0 e^{-{1\over3}K(z,
\ov z)}].
\eeq
The gauge choice for dilatations, which corresponds to impose
canonical Einstein terms, determines then $z_0\ov z_0$
[\ref{KU2}]:
\beq
\label{z0Cremmer}
z_0 \ov z_0= {1\over \kappa^2} e^{{1\over3}K(z, \ov z)}.
\eeq
This equation, together with the gauge fixing conditions on
the phase of $z_0$, $\psi_0$ and $b_\mu$ mentioned above,
are used in the superconformal theory to obtain the
super-Poincar\'e lagrangian.

The theory (\ref{Cremmer1}) is invariant under the K\"ahler
transformations
\beq
\label{Kahler1}
\begin{array}{rcl}
K &\longrightarrow& K + \varphi(\Sigma) + \ov\varphi(\ov\Sigma),
\\
w &\longrightarrow& e^{-\varphi(\Sigma)}w(\Sigma), \\
S_0 &\longrightarrow& e^{{1\over3}\varphi(\Sigma)}S_0,
\end{array}
\eeq
which can actually be used to eliminate the superpotential
(for $w\ne0$) by the choice $\varphi(\Sigma)=
\log w(\Sigma)$, equivalent
to a redefinition of the compensator defined by
\beq
\label{newcomp}
S_0^\prime = w(\Sigma)^{1/3}S_0,
\eeq
so that lagrangian (\ref{Cremmer1}) becomes
\beq
\label{Cremmer2}
{\cal L}= -{3\over2}\big[S_0^\prime\ov S_0^\prime
e^{-{1\over3}{\cal G}} \big]_D
+\big[{1\over4} f(\Sigma) W^AW^A  + S_0^{\prime\,3}\big]_F,
\eeq
where all couplings are now contained in the `K\"ahler
function' defined by
\beq
\label{KahlerG}
{\cal G}(\Sigma, \ov\Sigma e^V) = K(\Sigma, \ov\Sigma e^V)
+ \log |w(\Sigma)|^2.
\eeq
The equivalent form (\ref{Cremmer2}) shows that the
superpotential $w(\Sigma)$ and the K\"ahler potential
$K(\Sigma,\ov\Sigma e^V)$ only contribute to the lagrangian
in the combination ${\cal G}(\Sigma,\ov\Sigma e^V)$. Notice
that the gauge choice for dilatations is now
\beq
\label{Cremmercomp2}
z_0^\prime \ov z_0^\prime = {1\over\kappa^2} e^{{1\over3}
{\cal G}},
\eeq
which is compatible with eqs. (\ref{z0Cremmer}) and
(\ref{newcomp}).
The computation of the component expansion of lagrangian
(\ref{Cremmer2}), using the compensator (\ref{Cremmercomp2})
shows that it only depends on the functions $f(z)$ and
${\cal G}(z, \ov z)$ and on their derivatives.
Moreover, scalar kinetic terms have the simple form
$$
{\partial^2 {\cal G} \over \partial z\partial\ov z}
(\partial_\mu z)(\partial^\mu \ov z),
$$
which shows that $\cal G$ (or $K$) is the K\"ahler potential
for the non linear k\"ahlerian $\sigma$-model describing
scalar fields.

\subsection{Linear and chiral multiplets}

We now turn to the discussion of matter described by linear
and chiral multiplets. To simplify, we will only explicitly
consider one chiral multiplet $\Sigma$ and one real linear
multiplet $L$. The generalization to one linear multiplet
and an arbitrary number of chiral multiplets, which would
be relevant for string effective theories, is straightforward
at this point. We can freely choose the Weyl and chiral
weights for $\Sigma$ to be
$$
\Sigma: \qquad w = n = 0.
$$
On the contrary, a real linear multiplet has always
$$
L: \qquad w=2, \qquad n=0.
$$
However, since ${L\over S_0\ov S_0}$ is a real vector
multiplet with $w=n=0$, an invariant action for $\Sigma$
and $L$ would be
\beq
{\cal L}= \big[ S_0\ov S_0 \Phi({L\over S_0\ov S_0},
\Sigma, \ov\Sigma)\big]_D
+ \big[ S_0^3 w(\Sigma)\big ]_F.
\eeq
The introduction of gauge invariance, with a real vector
multiplet $V$, follows then the same principle as in
the case of global supersymmetry. One defines a vector
multiplet
\beq
\hat L = L - 2\Omega(V),
\eeq
$\Omega(V)$ being the Chern-Simons (vector) supermultiplet.
Gauge invariance of $\hat L$ is obtained by imposing the
gauge transformation
\beq
\delta L = 2\delta \Omega(V),
\eeq
an admissible condition since $\delta\Omega(V)$ is a linear
multiplet.
The general gauge invariant superconformal lagrangian for
$L$, $\Sigma$ and $V$ is then
\beq
\label{Llagran1}
{\cal L}= \big[ S_0\ov S_0 \Phi({{\hat L}\over S_0\ov S_0},
\Sigma, \ov\Sigma e^V)\big]_D
+ \big[ S_0^3 w(\Sigma)\big ]_F.
\eeq
It is important to remark that gauge kinetic terms do not
need to
be introduced separately, as in eq. (\ref{Cremmer1}) for
instance. They are contained in the Chern-Simons multiplet
$\Omega(V)$. The argument is essentially identical to the
case of global supersymmetry, and manipulations analogous
to (\ref{manip}) exist in the local context. Also, and for
the same reason as in global supersymmetry, a gauge kinetic
term $\big[ {1\over4}f(\Sigma)W^AW^A \big]_D$ can be transformed
into a lagrangian of the form (\ref{Llagran1}) so that this
general theory includes also arbitrary couplings for the
chiral multiplet $\Sigma$.

As before, a transformation
\beq
\label{ftransf}
\begin{array}{rcl}
w(\Sigma) &\longrightarrow& e^{-\varphi(\Sigma)}w(\Sigma),
\\
S_0 &\longrightarrow& e^{{1\over3}\varphi(\Sigma)}S_0 ,
\\
\Phi &\longrightarrow&e^{-{1\over3}[\varphi(\Sigma)+
\ov\varphi(\ov\Sigma)]}\Phi,
\end{array}
\eeq
can be used to eliminate the superpotential. The equivalent
theory has a new function $\Phi^\prime = [w(\Sigma)\ov
w(\ov\Sigma)]^{-1/3}\Phi$.

The linear multiplet with $w=2$ and $n=0$ is a real vector
multiplet with constrained components $(C, \chi, m, n, v_\mu,
\lambda, d)$. The conditions are
[\ref{KU}]
\beq
\label{Lcomp2}
m=n=0, \quad
D_\mu^c v^\mu =0, \quad
\lambda = - \gamma^\mu D_\mu^c \chi, \quad
d = -\Box^c C,
\eeq
where $D_\mu^c$ and $\Box^c$ are superconformal covariant
derivative and d'alembertian. Applying the density formula
(\ref{densities}) and tensor calculus to expression
(\ref{Llagran1}), one deduces that the lagrangian will
contain kinetic terms for the real scalar field $C$ and
Einstein terms of the form \footnote{
In the Wess-Zumino gauge, the lowest component of $\Omega$
vanishes.}
\beq
-e z_0\ov z_0{\partial\Phi\over\partial C} \Box^c C +
{1\over3}z_0\ov z_0\Phi eR,
\eeq
where it is understood that $\Phi$ is considered as a
function of the lowest components of $\Sigma$, $S_0$
and $\hat L$ only and not a full vector multiplet. Since
[\ref{PvN}]
\beq
\Box^c C = \Box C + {1\over3}RC + {\rm other\,\, terms}
\eeq
for a scalar field with Weyl weight two, the complete
Einstein term is
\beq
\label{Einstein1}
{1\over3}z_0\ov z_0 \left[ \Phi - C{\partial\Phi\over
\partial C} \right] eR.
\eeq
The gauge condition for dilatations which fixes $|z_0|$
is then
\beq
\label{Lcomp}
z_0\ov z_0 \left[ \Phi({C\over z_0\ov z_0}, z, \ov z) -
C{\partial\over\partial C} \Phi({C\over z_0\ov z_0}, z,
\ov z)\right] = -{3\over2} {1\over\kappa^2}.
\eeq
Contrary to the case of chiral multiplets only [see eq.
(\ref{z0Cremmer})], this is an implicit equation for
$z_0\ov z_0$ which appears in the arbitrary function
$\Phi$.

An elegant derivation of the Einstein term is as follows.
The lagrangian (\ref{Llagran1}) is equivalent to
\beq
\label{Llagran2}
{\cal L} = \big[ S_0\ov S_0 \Phi({U\over S_0\ov S_0}, \Sigma,
\ov\Sigma e^V) - (S+\ov S)(U + 2\Omega(V))\big]_D
+ \big[ S_0^3 w(\Sigma)\big ]_F,
\eeq
where $S$ is a chiral multiplet and $U$ an unconstrained
vector multiplet. The equations of motion for the components
of $S+\ov S$ simply impose that $U+2\Omega(V)$ is a linear
multiplet, hence the equivalence with eq. (\ref{Llagran1}).
In this lagrangian, the Einstein term is entirely due to
the D-density formula (\ref{densities}). It reads
$$
{1\over3}\left[ z_0\ov z_0 \Phi({u\over z_0\ov z_0}, z,
\ov z) - (s+\ov s)u \right]eR,
$$
where $s$ and $u$ are the lowest scalar components of $S$ and
$U$. In the Wess-Zumino gauge, $\Omega(V)$ does not contribute.
The equation of motion for $u$ is however
$$
z_0\ov z_0 {\partial\over\partial u} \Phi({u\over z_0\ov z_0},
z, \ov z)
= s+\ov s
$$
which leads to the Einstein term
$$
{1\over3} z_0\ov z_0\left[\Phi({u\over z_0\ov z_0}, z, \ov z) -
u {\partial\over\partial u}\Phi({u\over z_0\ov z_0}, z,
\ov z)\right]  eR,
$$
which is identical to expression (\ref{Einstein1}).

The form (\ref{Llagran2}) of the general lagrangian (\ref{Llagran1})
can be used to perform the duality transformation
which turns the linear multiplet $L$ into the chiral
one $S$. The procedure is similar to the case of
global supersymmetry. One solves the equation of motion
of the unconstrained vector multiplet $U$ which reads
\beq
\label{Ueomlocal}
\left[ {\partial\over\partial X}\Phi(X, \Sigma, \ov\Sigma e^V)
- (S+\ov S)\right]_{X = U (S_0\ov S_0)^{-1}} = 0,
\eeq
and performs
manipulations analogous to eqs. (\ref{manip}) to recast gauge
kinetic terms into an F-density involving the chiral multiplet
of gauge curvatures $W^A$. With the conventions we use
in our supergravity expressions,
$$
[(S+\ov S)\Omega(V)]_D = -{1\over2}[SW^AW^A]_F.
$$
The resulting lagrangian for
multiplets $S$, $\Sigma$ and $V$ is always of the form
\beq
\label{Slagranlocal}
{\cal L}_S = [S_0\ov S_0 {\cal H}(S+\ov S, \Sigma,
\ov\Sigma e^V)]_D
+[SW^AW^A + S_0^3w(\Sigma)]_F.
\eeq
Clearly,
\beq
{\cal H}(S+\ov S, \Sigma, \ov\Sigma e^V) =
\left [\Phi({U\over S_0\ov S_0}, \Sigma,
\ov\Sigma e^V) - (S+\ov S){U\over S_0\ov S_0 } \right],
\eeq
with ${U\over S_0\ov S_0}$ expressed as a function of $S+\ov S$,
$\Sigma$ and $\ov\Sigma e^V$ with the help of
equation of motion (\ref{Ueomlocal}). Theory
(\ref{Slagranlocal}) is of the form (\ref{Cremmer1}) with
an arbitrary function ${\cal H}$ but with
a universal gauge kinetic function
\beq
\label{fis4S}
f = 4 S.
\eeq
Notice that an analytic $S$-dependent field redefinition
of the compensator
\beq
\label{S0redef}
S_0 \longrightarrow g(S,\Sigma) S_0
\eeq
can be used to formally introduce a dependence on $S$
in the superpotential. One deduces easily that the most
general form of superpotential compatible with the
duality transformation is [\ref{CFV}]
\beq
\label{Ssuperpot}
w(S,\Sigma) = w_1(\Sigma) e^{\rho S}.
\eeq
with an arbitrary analytic function $w_1$ and a real number
$\rho$. As it should, this $S$-dependent superpotential
preserves the R--symmetry
$$
S \longrightarrow S + ia \qquad (a= {\rm real\,\,number})
$$
of the lagrangian (\ref{Slagranlocal}).

The component expansion of the general lagrangian
(\ref{Llagran1}) can be obtained by systematically using
tensor calculus. The presence of the Chern-Simons multiplet,
which is a complicated expression of the vector multiplet
$V$, and the implicit character of the compensator fixing
condition (\ref{Lcomp}) cause some technical difficulties. In
the following we will be interested in string effective
actions with known functions $\Phi$ and we will only consider
the scalar and gaugino bilinear contributions to the lagrangians.
This allows us to truncate the multiplets and the density formula,
disregarding unwanted components
and simplifying the computation of the component expansion.
For simplicity, we will also limit our results to a unique
gauge singlet chiral multiplet $\Sigma$ coupled to a
super-Yang-Mills multiplet for a simple gauge group which
will be identified with the hidden $E_8$
sector of $(2,2)$ heterotic strings, and to a supergravity
sector containing a linear multiplet. We will then use the
following truncated multiplets:
\begin{itemize}
\item{$\Sigma$:} we eliminate the fermionic components and
retain the scalars $z$ and auxiliary fields $f$.
\item{$L$:} the linear multiplet is a vector multiplet with
components (\ref{Lcomp2}). We omit its fermionic component
$\chi$ and the embedding of $L$ into a vector multiplet becomes
$c=C$, $B_\mu = v_\mu$, $d = -\Box^c C$, $\chi=m=n=\lambda=0$,
where $\Box^c$ is the superconformal covariant d'alembertian.
The constrained vector field $v_\mu$ can be expressed in terms
of an antisymmetric tensor using $v_\mu
={1\over\sqrt2}e^{-1}\epsilon_{\mu\nu\rho\sigma}\partial^\nu
b^{\rho\sigma}$, omitting gravitino terms.
\item{$\Omega$:} The Chern-Simons supermultiplet will be
the origin of gaugino bilinear terms. Its truncation is
embedded into a vector multiplet using $m+in = {1\over4}\lambda
\lambda$. All other components vanish.
\item{$V$:} The gauge vector multiplet appears explicitly
in the expression $e^V$ which does not contain any gaugino
bilinear term. It can then be truncated to zero.
\item{$S_0$:} We keep the scalar component $z_0$ and the
auxiliary field $f_0$.
\end{itemize}
To summarize, the embeddings into a vector multiplet
$(C, \chi, H, K, B_\mu, \Lambda, d)$ of the truncated
multiplets which will be used in the following are
\begin{equation}
\label{embed}
\begin{array}{rcl}
L &=& (C, 0, 0, 0, v_\mu, 0, -\Box^c C), \crbig
\Omega(V) &=& (0, 0, {1\over4}\ov\lambda\lambda, {i\over4}
\ov\lambda\gamma_5\lambda, 0, 0, 0) , \crbig
\Sigma &=& (z, 0, -f, if, iD_\mu^c z, 0, 0), \crbig
S_0 &=& (z_0, 0, -f_0, if_0, iD_\mu^c z_0, 0,0),
\end{array}
\eeq
where the Majorana four-component spinors $\lambda$
are the gauginos and a summation over all group generators
is understood. In the multiplet of superconformal gauge
fields, we will disregard all gravitino contributions.
This truncation implies also that the gauge field of
special supersymmetry, which is algebraic, can be omitted.
But the gauge field $A_\mu$ of chiral $U(1)$ rotations must
be kept: it is an auxiliary field of minimal Poincar\'e
supergravity and it contributes in particular to scalar
kinetic terms.
This truncation of the superconformal gauge fields leads to the simple
density formula already given in eqs. (\ref{densities}).

\nsect{Component lagrangians}

\subsection{Basics}

In preparation for the analysis of the effective supergravity
lagrangian of superstrings, we first apply the rules of superconformal
tensor calculus to the general lagrangian (\ref{Llagran1})
for one linear multiplet $L$ and one chiral multiplet
$\Sigma$. The truncated component expansion of this
theory before solving for auxiliary fields and fixing the
compensator $z_0$ is
\beq
\label{totalcomp}
{\cal L} =  {\cal L}_E + {\cal L}_{KIN} + {\cal L}_{AUX} +
{\cal L}_{4\lambda},
\eeq
where the four contributions give respectively the Einstein
term, the kinetic lagrangian of scalars and of the
antisymmetric tensor,
auxiliary fields terms, and quartic gaugino contributions.
As already mentioned [eq. (\ref{Einstein1}],
the Einstein term is
\beq
\label{LE}
e^{-1}{\cal L}_E = -{1\over2}R \left[ -{2\over3}\comp(\Phi -
C{\partial\over\partial C}\Phi) \right].
\eeq
Kinetic terms read:
\beq
\label{LKIN}
\begin{array}{rcl}
e^{-1}{\cal L}_{KIN} &=&
{1\over2} \Phi_{xx}\comp^{-1} (\partial_\mu C)
(\partial^\mu C)
-2\Phi_{z\ov z}\comp(\partial_\mu z)
(\partial^\mu \ov z) \crbig
&& - 2[\Phi-\Phi_x C\comp^{-1}+\Phi_{xx}C^2\comp^{-2}]
(\partial_\mu z_0)(\partial^\mu \ov z_0) \crbig
&&-2[\Phi_z-\Phi_{xz}C\comp^{-1}] z_0(\partial_\mu z)
(\partial^\mu \ov z_0) \crbig
&&-2[\Phi_{\ov z}-\Phi_{x\ov z}C\comp^{-1}] \ov z_0
(\partial_\mu \ov z)(\partial^\mu z_0) \crbig
&& -{1\over2}\Phi_{xx}\comp^{-1}v_\mu v^\mu - i
v_\mu(\Phi_{xz}\partial^\mu z - \Phi_{x\ov z}
\partial^\mu \ov z) \crbig
&& - \Phi_{xx} C\comp^{-1} iv_\mu \partial^\mu
\log\left({\ov z_0\over z_0}\right).
\end{array}
\eeq
The variable $x$ is the lowest component of the
multiplet $\hat L (S_0\ov S_0)^{-1}$, $x= C\comp^{-1}$,
and the notation for derivatives of $\Phi$ is
$$
\Phi_x={\partial\over\partial x} \Phi(x,z,\ov z), \qquad
\Phi_z= {\partial\over\partial z} \Phi(x,z,\ov z), \qquad
\ldots.
$$
Scalar and $b_{\mu\nu}$ kinetic terms
will receive further contributions from auxiliary
field $A_\mu$ when solving its
equation of motion. It actually turns out that after
solving for $A_\mu$, scalar
kinetic terms only depend on the compensator through the
combination $\comp$.
Fixing the compensator will then allow to express
$\partial_\mu \comp$ as a function of $C$, $z$, $\ov z$ and their
derivatives.
The part of the lagrangian involving auxiliary fields is
\beq
\label{LAUX}
\begin{array}{rcl}
e^{-1}{\cal L}_{AUX} &=&
2\left[ \Phi - \Phi_x C\comp^{-1} + \Phi_{xx} C^2\comp^{-2}
\right]
\left( f_0\ov f_0 - {1\over4}\comp A_\mu A^\mu\right) \crbig
&& +2[\Phi_z-\Phi_{xz} C\comp^{-1}]\left( z_0\ov f_0f
- {i\over2}\comp A_\mu\partial^\mu z\right) \crbig
&& +2[\Phi_{\ov z}-\Phi_{x\ov z} C\comp^{-1}]\left(
\ov z_0 f_0 \ov f + {i\over2}\comp A_\mu\partial^\mu \ov z\right)
\crbig
&&+ \left[ \Phi - \Phi_x C\comp^{-1} + \Phi_{xx} C^2\comp^{-2}
\right]iA_\mu
(z_0\partial^\mu \ov z_0 - \ov z_0\partial^\mu z_0) \crbig
&& + \Phi_{xx} C\comp^{-1} v_\mu A^\mu + 3z_0^2f_0w(z) +
3\ov z_0^2\ov f_0\ov w(\ov z)
+ z_0^3{dw\over dz}f + \ov z_0^3{d\ov w\over d\ov z} \ov f
\crbig
&& +\Phi_{xx} C\comp^{-2}\left[ z_0\ov f_0(\ov\lambda_L\lambda_R)
+ \ov z_0 f_0 (\ov\lambda_R\lambda_L) \right] \crbig
&& +2\Phi_{z\ov z} \comp f\ov f -
\Phi_{xz}f(\ov\lambda_R\lambda_L)
-\Phi_{x\ov z}\ov f (\ov\lambda_L\lambda_R).
\end{array}
\eeq
A summation on all gaugino spinors is understood in gauge
invariant bilinear expressions like $(\ov\lambda_L\lambda_R)$.
The chiral auxiliary fields $f$ and $f_0$
only produce potential
and non-derivative gaugino terms.
Finally, the last contribution to (\ref{totalcomp}) is a quartic
gau\-gino contribution:
\beq
\label{Lfourlambda}
e^{-1}{\cal L}_{4\lambda} = {1\over2} \Phi_{xx}
\comp^{-1}(\ov\lambda_L\lambda_R)(\ov\lambda_R\lambda_L).
\eeq
The auxiliary fields of minimal Poincar\'e supergravity are $A_\mu$
[chiral $U(1)$ gauge field] and $f_0$ [in the chiral
compensator $S_0$]. The Poincar\'e theory will be obtained by
solving for $A_\mu$ and $f_0$, and by fixing the compensator with the
requirement
$$
-{2\over3}\comp[\Phi - \Phi_x C \comp^{-1}] = {1\over\kappa^2}\,,
$$
which canonically normalizes the Einstein lagrangian.

Using the Weyl weights
$$
\begin{array}{rrclrcl}
L:& \quad\quad w(C)&=&2 \quad&\quad w(v_\mu)&=& 3 \\
\Sigma:& \quad\quad  w(z)&=&0 \quad&\quad w(f)&=&1 \\
S_0:& \quad\quad w(z_0)&=&1 \quad&\quad w(f_0)&=&2 \\
&\quad w(A_\mu)&=&1 \quad&&&
\end{array}
$$
which specify the physical dimensions of
the various component fields,
one easily checks that every term in the lagrangian (\ref{totalcomp})
has dimension four.

\subsection{K\"ahler invariant lagrangians in components}

As mentioned above, the component lagrangian (\ref{totalcomp})
can be used to construct the scalar and gaugino bilinear sector of
a general $N=1$ supergravity with a linear multiplet.
The general form is actually not
very illuminating and, moreover, the equation (\ref{Lcomp})
used to determine the compensator can only be solved
implicitly for an arbitrary $\Phi$.
We will then concentrate on a specific class of functions $\Phi$
which is of direct interest in the context of the effective
supergravity theory of superstrings.

Supergravity couplings of chiral multiplets are characterized by
K\"ahler invariance (\ref{Kahler1}).
It is also known that superstrings often possess symmetries
which act in the effective supergravity like K\"ahler
symmetries. This means that there exists
a K\"ahler potential $K$ which, together with the superpotential
$w(\Sigma)$ and the compensator $S_0$, transforms according to
eqs. (\ref{Kahler1}). An example of such a symmetry of stringy
origin is target-space duality [\ref{duality}].
The K\"ahler potential, which is in
general a gauge invariant function of $\Sigma$ and $\ov\Sigma e^V$,
can be regarded as a composite connection for K\"ahler symmetry. The
combination
$$
{e^{K/3}\hat L\over S_0\ov S_0}
$$
is a K\"ahler and gauge invariant quantity with conformal weight zero.
The class of functions $\Phi$ of the form
\beq
\label{useless1}
\Phi=\frac{\hat{L}}{S_0 \ov{S}_0} {{F}}
\left( e^{K/3} \frac{\hat{L}}{S_0 \ov{S}_0}\right),
\eeq
with ${{F}}$ an arbitrary real function, leads then to K\"ahler invariant
superconformal theories with a non-trivial coupling of the linear multiplet.
Notice that this choice is not the most general one since a further
dependence on gauge and K\"ahler invariant
functions of the chiral multiplet is in
principle allowed, a possibility which will
not be considered here. Notice also that the tree-level
effective supergravity of superstrings is of the
form (\ref{useless1}) with [\ref{CFV}]
$$
F(y) = -{1\over\sqrt2}\, y^{-3/2}.
$$
The advantage of (\ref{useless1}) is that
the K\"ahler potential $K(\S,\ov \S e^V)$, which controls the standard
complex geometry of the chiral superfields, is explicit in $\Phi$.
The function $\Phi(z,\ov z)$ appears in the component
lagrangian as a connection which, for instance,
changes $\comp$ to the K\"ahler invariant
expression
$z_0e^{-K/3}\ov z_0$. We will later verify that the scalar kinetic
terms are as usual given by the second derivatives of $K$,
as it should for a K\"ahler potential.

Let us write the auxiliary-fields lagrangian (\ref{LAUX}) as:
\beq
\label{LAUXD}
\begin{array}{rcl}
e^{-1}{\cal L}_{AUX}&=&{\c A} f_0\ov f_0 +\left({\c B} \ov f_0 f+{\rm h.c.}
\right)
+{\c C} f\ov f -\frac{1}{4}\comp {\c A} A_\mu A^{\mu}\crbig
&& + {\c D} v_\mu A^\mu+\left({\c E}f_0
+{\c F} f-\frac{i}{2}\ov z_0 {\c B}\partial_\mu z A^\mu +
{\rm h.c.}\right)\crbig
&&+{i\over2} z_0\ov z_0 {\c A} A_\mu \partial^\mu\log (\ov z_0/z_0) ,
\end{array}
\eeq
where the coefficients can be read
from (\ref{LAUX}) and for the case (\ref{useless1})
are given by:
\beq
\label{ABC}
\begin{array}{rcl}
{\c A}&=&2[\Phi-x\Phi_x+x^2\Phi_{xx}]=2 x y
(F'+yF''), \crbig
{\c B}&=&2 z_0[\Phi_z-x\Phi_{xz}]=-\frac{2}{3}
z_0 x y (F'+y F'')K_z, \crbig
{\c C}&=&2\comp \Phi_{z\ov z}=\frac{2}{9}\comp x y
\left[3 F' K_{z \ov z}+(F'+ yF'')K_z K_{\ov z}\right], \crbig
{\c D}&=&x\Phi_{xx}=y(2F'+ yF''), \crbig
{\c E}&=&3z_0^2w + z_0^{-1} x\Phi_{xx}(\ov\lambda_R\lambda_L)=
3z_0^2w(z)+  z_0^{-1}  xy(2F'+yF'')(\ov\lambda_R\lambda_L), \crbig
{\c F}&=&z_0^3 w_z -\Phi_{xz}(\ov\lambda_R\lambda_L)=
z_0^3 w_z-\frac{1}{3}y(2F'+yF'')(\ov\lambda_R\lambda_L) K_z .
\end{array}
\eeq
We use again $x = C\comp^{-1}$ and $F', F''$ refer to derivatives
of $F$ with respect to its argument $y\equiv xe^{K/3}$.
In (\ref{LAUXD}), the last coupling of the form
$A_\mu\partial^\mu \log(\ov z_0/z_0)$ can be eliminated
using the fact that the compensator will
be fixed to be real. To compute the scalar
potential we have to solve the equations for $f$
and $f_0$. They give:
\beq
\label{SOL}
\begin{array}{rcl}
f_0&=&({\c A}{\c C}-{\c B}\ov{\c B})^{-1} \left({\c B}\ \ov{\c F}-
{\c C}
\ \ov {\c E}\right), \crbig
f&=&({\c A}{\c C}-{\c B}\ov{\c B})^{-1} \left(\ov{\c B}\ \ov{\c E}-{\c A}
\ \ov{\c F}\right) ,
\end{array}
\eeq
and the potential is obtained by inserting these expressions
into the lagrangian. It is in general given by
\beq
\label{POT}
\begin{array}{rcl}
V_{aux}&=&-(\ov{\c E}\ \ov f_0 +\ov{\c F}\ \ov f)\crbig
&=&({\c A}{\c C}-{\c B}\ov{\c B})^{-1}\left[{\c C}|{\c E}|^2+{\c A}
|{\c F}|^2
-(\ov{\c E}\ov{\c B} {\c F}+ {\rm h.c.})\right] .
\end{array}
\eeq
Using eqs. (\ref{ABC}), we can obtain the scalar potential
including the contribution generated by expectation
values of gaugino bilinears, provided we
also add the four--gaugino term (\ref{Lfourlambda}). We obtain
\beq
\label{POTD}
\begin{array}{rcl}
V&=&V_{aux}+V_{4\lambda}\crbig
&=&{3\over2}\comp^2 x^{-2}e^{-K/3} F'^{-1} K_{z\ov z}^{-1}
\left\vert w_z+wK_z\right\vert ^2\crbig
&& +{9\over2}\comp^{-1} x^{-2}(F'+ yF'')^{-1}e^{-K/3}\left
\vert z_0^3w+\frac{1}{3}y(2F'+yF'')
(\ov\lambda_R\lambda_L)\right\vert ^2\crbig
&&-{1\over2}\comp^{-1} e^{K/3}(2F'+yF'')\left\vert
(\ov\lambda_R\lambda_L)
\right\vert ^2 ,
\end{array}
\eeq
where the last term is the contribution from $V_{4\lambda}$.

At this level of generality, our analysis extends trivially
to couplings of one linear multiplet $L$ and any number of
chiral multiplets $\S_i$ to supergravity.
In this case the coefficient ${\c C}$ is actually
a matrix
${\c C}_{i\ov{\ell}}$ and ${\c B}$ and ${\c F}$ are vectors
${\c B}_i$ and ${\c F}_i$ respectively. The auxiliary field
$f$ is also a vector $f_i$. Then the expression in the scalar
 potential $K_{z\ov z}^{-1}|w_z+wK_z|^2$ should be seen as
a particular case of the general expression
$K_{i\ov{\ell}}^{-1} (w_i+wK_i)(\ov w_{\ov{\ell}}+\ov w K_{\ov{\ell}})$.
It is a nontrivial fact that the matrix
$({\c A}{\c C}_{i\ov{\ell}}-{\c B}_{i}\ov{\c B}_{\ov{\ell}})^{-1}$
leads to a simple expression
only
in terms of $K_{i\ov{\ell}}^{-1}$. Having mentioned this, for
simplicity, we will continue writing the expressions in terms of only
 one
chiral superfield.

To obtain the scalar kinetic terms we need first to solve the field
equation for the auxiliary field
$A_\mu$. From (\ref{LAUXD}), we can see that
\beq
\label{Amu}
\begin{array}{rcl}
A_\mu &=& 2\comp^{-1} {\c A}^{-1} \left( {\c D}v_\mu -{i\over2}
[\ov z_0{\c B}\partial_\mu z - z_0\ov{\c B}\partial_\mu\ov z]
\right) +i\partial_\mu
\log \left({\ov z_0\over z_0}\right) \crbig
&=&  {i\over3}[K_z\partial_\mu z-K_{\ov z}\partial_\mu \ov z]
+ \comp^{-1}x^{-1} {2F'+yF'' \over F'+yF''} v_\mu
+i \partial_\mu \log\left( {\ov z_0\over z_0}\right).
\end{array}
\eeq
The $A_\mu$-dependent terms in (\ref{LAUXD}) will then add contributions
of the form
$$
e^{-1} \Delta {\c L}_{KIN} = {1\over4}\comp{\c A} A_\mu A^\mu
$$
to scalar kinetic terms (\ref{LKIN}). It turns out that the
complete kinetic terms
can be written in an elegant form before fixing the compensator:
\beq
\label{KINbefore}
\begin{array}{rcl}
e^{-1}{\cal L}_{KIN} &=&
-{1\over2}\comp^{-1}e^{K/3}\,
{2F'+yF'' \over F'+yF''} F'\left[ (\partial_\mu C)(\partial^\mu C)
-v_\mu v^\mu \right] \crbig
&& -{2\over3} \comp xy F' K_{z \ov z} (\partial_\mu \ov z)
(\partial^\mu z) \crbig
&& +\comp xy {F'\over F'+yF''}\left[ (2F'+yF'') C^{-1}(\partial_\mu C)
(\partial^\mu\Delta) - {1\over2} F' (\partial_\mu \Delta)
(\partial^\mu\Delta) \right],
\end{array}
\eeq
where $\Delta$ is the combination
$$
\Delta=\log [z_0\ov z_0 xyF'] = \log[ \comp^{-1}C^2 e^{K/3} F'],
$$
which will be equal to a constant with the compensator fixing
condition (\ref{Lcomp}), which in our case reads
\beq
\label{EinsteinF}
\comp ^{-1} C^2 e^{K/3} F' = e^\Delta= {3\over2}{1\over\kappa^2}.
\eeq
This condition can be explicitly solved only after having specified
the function $F$. However, for an arbitrary $F$, scalar and
antisymmetric tensor kinetic terms will become
\beq
\label{LKINPoincare}
e^{-1}{\cal L}_{kin.}= -{1\over\kappa^2}K_{z\ov z}
(\partial_\mu \ov z)(\partial^\mu z)
- {3\over4}{1\over\kappa^2}
\frac{2F'+yF''}{F'+yF''} C^{-2} \left[
(\partial_\mu C)(\partial^\mu C)-v_\mu v^\mu\right]
\eeq
in the Poincar\'e supergravity theory.
This shows that the K\"ahler potential $K$, introduced first as a
connection to construct K\"ahler invariant functions, does give
the $\sigma$--model metric for the chiral scalar fields, as it should.
Notice also that since the ratio ${(2F'+yF'')}/{(F'+yF'')}$
must be positive to ensure positivity of kinetic energy, we can see
that the last two  terms in the scalar potential (\ref{POTD})
have opposite signs.  The potential is not explicitly
positive definite.

A nontrivial check for all these expressions is
to take the
particular case $F(y)=-\frac{3}{2y}$ which corresponds to
$\Phi=-\frac{3}{2}e^{-K/3}$, the general form (\ref{Cremmer1}) for chiral
multiplets coupled to
supergravity only. This choice of $F$
implies that $2F'+yF''=0$, and equations (\ref{KINbefore}),
(\ref{EinsteinF}) and (\ref{LKINPoincare}) become respectively
$$
\begin{array}{rcl}
e^{-1}{\cal L}_{KIN} &=& -\comp e^{-K/3}K_{z\ov z}
(\partial_\mu \ov z)(\partial^\mu z) \crbig
&&+{3\over4}\comp e^{-K/3} (\partial_\mu\log z_0\ov z_0 e^{-K/3})
(\partial^\mu\log z_0\ov z_0 e^{-K/3}), \crbig
\comp e^{-K/3} &=& {1\over\kappa^2}, \crbig
e^{-1}{\cal L}_{kin.} &=& -{1\over\kappa^2}K_{z\ov z}
(\partial_\mu \ov z)(\partial^\mu z).
\end{array}
$$
The scalar potential (\ref{POTD}) reduces to
\beq
\kappa^4 V(z,\ov z) =e^K\left(
 K_{z\ov z}^{-1}\left\vert w_z+K_z w\right\vert ^2-
3\left\vert w\right\vert ^2\right)
\eeq
which is the well-known potential obtained in ref.
[\ref{CFGVP}]. In some sense, our expressions are more general
since they include the general couplings of a linear multiplet.
The effective supergravity of superstrings is, up to some subtleties,
a particular class of the models discussed in this section.
To this we will turn next.

\nsect{Superstring effective actions and scales}

We now turn to the study of the effective lagrangians
which describe at low energies the physics
of the four-dimensional heterotic superstring theory, with
a particular attention to the problem of scales in the effective
supergravity theory.
While there is no ambiguity at the string tree-level,
the derivation of an effective theory is more subtle
when loop corrections of string origin are also considered.
At string tree-level, the definition of a low-energy effective
action is simple since small external momenta imply small momenta
also on internal lines of tree-level diagrams. The problem
requires more attention when loop
diagrams are taken into account at the
string and effective field theory levels.

A string loop calculation, like the gauge coupling constants
calculations performed in [\ref{DKL}, \ref{ANT}], requires
the introduction of an infra-red cut-off $M_{IR}$ which
avoids divergences when massless string states circulate
on loops with small momenta. The cut-off forbids
momenta smaller than $M_{IR}$.
Since $M_{IR}$ is an arbitrary scale, its variation should
reflect in changes of physical quantities
controlled by renormalization-group equations.
For instance, it
has been verified that one-loop gauge coupling constants
computed in strings show the dependence on $\log M_{IR}$
implied by the one-loop gauge $\beta$-function for massless
string modes [\ref{M}, \ref{K}].

What we are interested in is the description of the
dynamics of string massless modes using an effective
low-energy field theory. This can be done for energies
below a physical ultraviolet cut-off, $M_{UV}$.
This quantity is arbitrary, but it should not exceed
the order of magnitude of the
mass of the lightest string excited modes.
Physical quantities computed in the effective theory,
which depend on $M_{UV}$, should correspond to the same
quantities computed in the string theory, which depend
on stringy scales (string tension and compactification radii)
but also on $M_{IR}$. The infra-red cut-off $M_{IR}$ may be
identified with $M_{UV}$, but this is not necessary: in general,
the effective field theory will also depend explicitly on
the scale $M_{IR}$, and it should be understood as an effective
theory in the Wilson sense.

In the string context,
the {\it Wilson effective action} $S_W$ is a local field
theory obtained by
integrating out all heavy string and Kaluza-Klein modes, at the
loop level, with the prescription that
one integrates all loops with virtual
momenta larger than a scale $\mu$ which characterizes
the Wilson action. The `small' scale $\mu$
acts like an infra-red cut-off and it should be identified with
the string infra-red cut-off $M_{IR}$.
The Wilson action has a formal expansion
in string loops, but a string amplitude for massless external states
at $n$ string loops is computed by summing all diagrams of order $n$
constructed from terms in the Wilson action up to
order $n$. For instance, a
one-loop string amplitude in the formalism of the Wilson action
combines two contributions:
one-loop diagrams from the Wilson tree-level action, which correspond
to massless string modes on the loop, and
tree diagrams from the one-loop correction to the Wilson action,
corresponding to the effect of heavy string modes
and virtual momenta larger than $\mu$
in the string loop.

A second possibility would be to use a generalisation of the
{\it effective action} $S_\Gamma$ of quantum field theory,
which is the
generating functional of one-particle irreducible Green's functions.
In the string context, $S_\Gamma$ would be the generating
functional of
string amplitudes for external massless states.
Contrary to the Wilson action, it also includes the
contributions of light or massless loops.
With massless states, the effective action is a non-local
functional of external classical fields.
The relation between $S_W$ and $S_\Gamma$ is very simple:
$S_\Gamma$ is the effective action, as defined in field theory,
for the field theory with action $S_W$.
To define the effective action $\SG$, we need an ultraviolet
cut-off which we will identify with $z_0\ov{z}_0$ and an arbitrary
running scale $\mu$ which can be varied by the action of the renormalization
group. The arbitrary scale $\mu$ corresponds to a subtraction point, to
some choice of
momenta of external gauge bosons in an amplitude used to normalize the
gauge coupling constant.
The invariance of $\SG$ under a change in $\mu$ is realized
by the fact that physical quantities satisfy
renormalization-group equations.

A symmetry of loop-corrected string amplitudes, like for
instance target-space duality, is also a symmetry of the
effective action. However, it is not necessarily a symmetry
of the Wilson action. The Wilson action can be anomalous, with
the perturbative
anomaly cancelled in amplitudes by a mechanism analogous to
gauge and gravitational anomaly cancellation in ten-dimensional
heterotic strings [\ref{GS}]. This phenomenon has been established
for target-space duality in $(2,2)$ superstrings,
in the sector of gauge kinetic terms [\ref{DFKZ1}, \ref{DKL},
\ref{ANT}]. It turns out that target-space duality acts in
the tree-level Wilson action like an anomalous
K\"ahler symmetry with the anomaly cancelled by quantum corrections to
$S_W$, a mechanism which has been studied in general terms in the
context of supergravity theories [\ref{CO}, \ref{CO2}].

In the following discussion, we will
use the terminology {\it effective theory} for a local
field theory corresponding in fact to the Wilson approach to
effective lagrangians. We will explicitly mention when we will consider
quantities related to the {\it effective action $S_\Gamma$.}

\subsection{String tree-level effective actions and
non renormalization theorems}

We consider now the $d=4$, $N=1$ heterotic string effective actions.
We will mainly focus in this section on general information that can
be obtained from the superfield formulation leaving
the explicit component expressions for the
next section.
For simplicity, whenever our expressions refer explicitly
to string actions we will restrict to the
diagonal overall modulus $T$, a chiral superfield
which exists in $(2,2)$ symmetric orbifolds
(as well as in $(2,2)$ Calabi-Yau compactifications).
The set of all chiral supermultiplets, denoted collectively
by $\Sigma$, will then contain $T$ as well as (charged) chiral matter
multiplets $Q_I$.
Later on, when discussing loop corrections,
we will also concentrate on the $(2,2)$ $Z_3$ and
$Z_7$ symmetric orbifolds for which the threshold
corrections to the gauge
coupling constants are known to vanish [\ref{DKL}].

As mentioned before, the dilaton in string theory
is part of a linear multiplet $L$ which includes the antisymmetric
tensor
field and a Majorana fermion. The most general lagrangian
for supergravity coupled to one linear multiplet  and any number
of chiral multiplets $\S$ depends on two arbitrary functions,
a real function  $\Phi(L-2\Omega,\S,\bar\S e^V)$ and
the superpotential $w(\S)$, which can be eliminated
by a transformation of the form (\ref{ftransf}).
It is convenient to express the
full action as a superconformal density [see (\ref{Llagran1})]:
\beq
\label{lgen}
{\cal L}= \left[
S_0 \ov{S}_0 \Phi \left(\frac{\hat{L}}{S_0 \ov{S}_0},\S,\ov{\S} e^V
\right)
\right]_D
+\left[S_0^3 w \right]_F ,
\eeq
where $\hat{L} = L - 2\Omega \equiv L - 2\sum_a \Omega_a$, $\Omega_a$
is the Chern-Simons
superfield associated with the gauge group factor $G_a$.
As usual, $S_0$ is
the compensating chiral multiplet of conformal supergravity which will be
fixed to give canonical Einstein kinetic term in the component action
of the Poincar\'e supergravity.
Notice that unlike the pure
chiral superfield case [\ref{CFGVP}] the gauge coupling is
{\it not} an independent arbitrary function but comes from $\Phi$ through
the $\Omega$ dependence. For the case of superconformal densities
defined in (\ref{lgen}), it is actually given by
\beq
\label{gauge}
\frac{1}{g^2}= 2z_0 \ov{z}_0 \frac{\partial \Phi}{\partial C},
\eeq
where $z_0\ov z_0 {\partial \Phi\over\partial C}$
is the lowest component of the multiplet $S_0\ov S_0
{\partial\Phi\over\partial
L}$. This result follows from the component expansion
of $\Phi$ and can be easily verified for instance from eq.
(\ref{fis4S}) and the duality transformation (\ref{Ueomlocal}).
At string tree-level, $\Phi$ is given by [\ref{CFV}]
\beq
\label{phi}
\Phi_{\rm 0}=-\frac{1}{\sqrt 2}\left(
\frac{S_0 \ov{S}_0}{\hat{L}}\right)^{1/2} e^{-K/2}
=
-{1\over\sqrt2}{\hat L\over S_0\ov S_0} \left(
{\hat L\over S_0\ov S_0}e^{K/3}\right)
^{-3/2},
\eeq
which belongs to the class of K\"ahler invariant theories
(\ref{useless1}). In the tree-level lagrangian
\beq
\label{treelag}
{\cal L}_{\rm 0}=-\frac{1}{\sqrt 2} \left[
\left( S_0 \ov{S}_0\right)^{3/2} {\hat L}^{-1/2} e^{-K/2}
\right]_D + \left[S_0^3 w \right]_F ,
\eeq
we find the conformal and K\"ahler invariant gauge coupling
\beq
\label{gbare}
\frac{1}{g^2(z_0\ov{z}_0)}=U\equiv 2\left(\frac{z_0
\ov{z}_0 }{2{C} e^{K/3}}\right)^{3/2} ,
\eeq
where $z_0$ and $C$ are the first components of $S_0$ and $L$
respectively
\footnote{Hereafter, whenever couplings are related with fields,
as in (\ref{gbare}), it should be understood as a relation for
the
vacuum expectation values of the fields.}.
We explicitly indicate that this gauge coupling depends on
$\comp$ since we have not yet fixed the conformal invariance.

We now wish to discuss in more detail the
choice of the compensator field $\comp$ and the r\^ole of the various
fields at string tree-level. For this purpose, we consider
the tree-level kinetic terms for the graviton, the gauge fields
and the antisymmetric tensor. Using (\ref{phi}), these terms
read
\beq
\label{treekin}
e^{-1} {\cal L}_{0,kin} = {1\over\sqrt2}[z_0\ov z_0 e^{-K/3}]
^{3/2} C^{-1/2} \left[ -{1\over2}R - {1\over4}C^{-1} F_{\mu\nu}^A
F^{A\,\mu\nu} +{1\over4}C^{-2} v_\mu v^\mu \right].
\eeq
Since $v_\mu = {1\over\sqrt2}e^{-1}\epsilon_{\mu\nu\rho\sigma}
\partial^\nu b^{\rho\sigma}$, the last term is a kinetic lagrangian
for the antisymmetric tensor,
with
$$
v^\mu v_\mu = -3 H^{\mu\nu\rho}H_{\mu\nu\rho} ,
\qquad
H_{\mu\nu\rho} = {1\over3}\left (
\partial_\mu b_{\nu\rho} + \partial_\nu b_{\rho\mu}
+ \partial_\rho b_{\mu\nu}\right).
$$
The dual theory, with $C$ replaced by the real field $s+\ov s$ is obtained
using the procedure described in section 3 [see eqs. (\ref{Llagran2}),
(\ref{Ueomlocal}) and (\ref{Slagranlocal})]. Since
\beq
\label{stree}
s + \ov s = \left({ z_0\ov z_0 \over 2Ce^{K/3}}\right)^{3/2},
\eeq
the dual kinetic terms are
\beq
\label{streekin}
\begin{array}{rcl}
e^{-1} {\cal L}_{0,kin} &=&
-{1\over 2} z_0\ov z_0(s+\ov s)^{1/3} e^{-K/3}\, R \crbig
&&-{1\over2} (s+\ov s) F_{\mu\nu}^A F^{A\,\mu\nu}
+ (s+\ov s)^2 [z_0\ov z_0(s+\ov s)^{1/3} e^{-K/3}]^{-1}
v_\mu v^\mu.
\end{array}
\eeq
Notice that
the Einstein term is the same as in eq. (\ref{ECremmer}),
with the replacement $K \longrightarrow
 -\log(s+\ov s)+K$.
Suppose that we fix dilatation symmetry by choosing the compensator
$\comp$ such that the Einstein term is canonical \footnote{
This choice corresponds to the `Einstein frame'.
}. The condition is
\beq
\label{EFcomp}
{1\over\sqrt2}[z_0\ov z_0 e^{-K/3}]^{3/2} C^{-1/2} =
UC = {1\over\kappa^2},
\eeq
and the kinetic terms become
\beq
e^{-1}{\cal L}_{0,kin} =
-{1\over2\kappa^2}R - {1\over4}(\kappa^2C)^{-1} F_{\mu\nu}^A
F^{A\,\mu\nu} + {1\over4}\kappa^2 (\kappa^2C)^{-2} v_\mu v^\mu.
\eeq
The tree-level gauge coupling constant in the Poincar\'e  theory
is simply
\beq
\label{g2Poin}
{1\over g^2} = \langle{1\over\kappa^2 C}\rangle,
\qquad
{1\over\kappa^2 C} = 2(s+\ov s).
\eeq
Instead, we could prefer to fix dilatation symmetry to obtain
the `string frame', with kinetic terms of the form [\ref{DS1}]
$$
{\varphi^{-1/2}\over\kappa^2}
\left[ -{1\over2}R - {1\over4}M_s^{-2} F_{\mu\nu}^A
F^{A\,\mu\nu} +{1\over4}M_s^{-4} v_\mu v^\mu \right].
$$
The real scalar field $\varphi$ is the string `loop-counting parameter'
and the string scale $M_s$ is related to the gauge coupling constant
by $M_s = g/\kappa$.
Clearly, the `string frame' corresponds to the condition
\beq
\label{STRcomp}
{1\over\sqrt2}[z_0\ov z_0 e^{-K/3}]^{3/2} = {1\over\kappa^3},
\eeq
instead of (\ref{EFcomp}). The comparison of kinetic terms shows that
$\varphi=\kappa^2C$ is the string loop-counting parameter, that
$$
M_s^2 = \langle C \rangle,
$$
while the gauge coupling constant is again $g^2= \kappa^2\langle C
\rangle$.
But, according to (\ref{stree}) and (\ref{g2Poin}), the field $s +\ov s$
of the dual theory is given by
$$
s + \ov s = {1\over2}\left({1\over\kappa^2C}\right)^{3/2},
$$
instead of the second equality (\ref{g2Poin}).

The $\S$ dependence of the tree-level lagrangian (\ref{treelag})
is encoded in the K\"ahler potential only:
\beq
\label{kahler}
K(T,\ov{T},Q,\ov{Q})= -3\log (T+\ov{T}) + \sum_I
(T+\ov{T})^{n_I}Q_I \bar Q_I
+\ldots
\eeq
and the superpotential $w(Q)=Y_{IJK}(T)Q^IQ^JQ^K+\ldots$
where $n_I$ are different `modular weights' associated to the respective
matter fields, $Y_{IJK}(T)$ the modulus dependent Yukawa couplings
and the ellipsis refer to higher order terms in the $Q^I$ expansion.
Because of K\"ahler invariance, the theory depends only on
the single function ${\cal{G}}= K+\log{|w|}^2$. This can be seen
by redefining the compensator, $S_0^{'3}=S_0^3 w$, with a transformation
similar to (\ref{ftransf}),
\beq
\label{ktransf}
\left\{
\begin{array}{lcl}
K & \rightarrow & K + \varphi + \ov{\varphi} \\
S_0 & \rightarrow & S_0 e^{\varphi/3} \\
w & \rightarrow & w e^{- \varphi}
\end{array}
\right. ,
\eeq
which leaves $\hat L$ invariant.
Target-space duality is an example of a symmetry which
manifests itself as a  K\"ahler transformation. Its action on $T$ is
such that $\cal G$ is invariant, but $K$ and $w$ transform as
in (\ref{ktransf}).

In this formalism, the linear supermultiplet $L$, which
contains the dilaton field related  to the string coupling
as its lowest component, has by supersymmetry a nature different
from the other multiplets $T$ and $Q_I$.
This observation allows to extract
very useful information about the possible loop corrections to the
effective lagrangian.
First we can say that since the superpotential cannot depend on the
field $L$,  it cannot get any
quantum correction, perturbative or non--perturbative, as long
as this formalism is applicable. This is a strong result that goes
beyond previous arguments which apply only at the perturbative level and
were obtained in the dual formalism where $L$ is transformed into
the chiral multiplet $S$ \footnote{
This argument does not imply that the superpotential
in the dual formalism is $S$-independent. See section 3.2,
and eq. (\ref{Ssuperpot}).
}. In this approach, it was shown that
because of a Peccei--Quinn (PQ)
symmetry under which the imaginary part of this field is shifted,
$S$ cannot appear in the superpotential
[\ref{BFQ}]. Showing that this symmetry is preserved in
perturbation theory gave the  non-renormalization theorem
of [\ref{DS}]. We have already seen that the existence of
this symmetry acting on $S$ is a consequence of the duality
equivalence of $S$ with the linear multiplet $L$ for which the
PQ symmetry is hidden in the gauge symmetry acting on
the antisymmetric tensor $b_{\mu\nu}$.
Working with the linear multiplet implies that the PQ symmetry
is exact and, as long as perturbative or non-perturbative corrections
can be expressed in this formalism, the PQ symmetry will remain
unbroken.

A similar comment can be made about the K\"ahler potential which,
being defined as a function of
chiral superfields only, cannot get any loop
corrections.
This result is however not as powerful as it sounds.
Besides the K\"ahler potential $K$ itself,
the main information in the function $\Phi$ is the coupling of
$L$ to the chiral multiplets described by $K$, a coupling
which is left entirely unconstrained. One could try to invoke
K\"ahler symmetry under transformations (\ref{ktransf})
to restrict this arbitrariness, to argue for instance that
$\Phi$ should be of the form
\beq
\label{useless2}
\Phi=\frac{\hat{L}}{S_0 \ov{S}_0} {\cal{F}}
\left( e^{K/3} \frac{\hat{L}}{S_0 \ov{S}_0}\right),
\eeq
with ${\cal{F}}$ an arbitrary real function.
The actual situation is however more complicated
since K\"ahler invariance is related to quantum string symmetries
of the effective action $S_\Gamma$, which are
plausibly anomalous when considered
using Wilson's effective lagrangian. This is at least the case of
target-space duality in $(2,2)$ theories. Such symmetries of the
effective action correspond to a
Green-Schwarz anomaly-cancellation mechanism which introduces in
the Wilson lagrangian counterterms which {\it do not}
preserve the
symmetries, as we will see below. Equation (\ref{useless2}) applies only
to the tree-level, invariant terms.
As far as $\Phi$ is concerned, the existence of
exact symmetries, acting like K\"ahler symmetries, and realized in
an anomaly-cancellation mode, tells us that $\Phi$ cannot be of the form
(\ref{useless2}) since the Wilson lagrangian is not invariant.

\subsection{Loop-corrected effective actions}

Very powerful information concerning the general form
of the loop-corrected effective lagrangians can be obtained by using
the fact that the gauge
coupling constant is given by equation (\ref{gauge}). In superstrings,
the gauge coupling constant is the expectation value of
a function of scalar fields. Then, loop corrections to gauge couplings
in the form of renormalization-group equations should provide
information of the functional dependence on $L$ of the function $\Phi$
defining the loop-corrected effective theory.
Fortunately, loop corrections to gauge couplings in strings
is a subject for which
calculations have been performed in large classes of superstrings
[\ref{DKL}, \ref{ANT}].  The
different contributions to the full string loop amplitude from the massless
and massive sectors are well understood, allowing then
to know the corrections to the effective gauge coupling at
energies below the Planck scale.
To discuss the loop corrections to the effective gauge couplings we need to
differentiate between the Wilson action ($S_W$) and the
one--particle--irreducible effective action $\SG$ as mentioned above.
For making our discussion self-contained, we will reproduce here the
one-loop superfield effective action for the simplest orbifold models,
namely $(2,2)$ $Z_3$ and $Z_7$ orbifolds.

The threshold corrections to the gauge couplings were computed
in [\ref{K}, \ref{DKL}] for $(2,2)$ orbifold models performing
a one-loop string
calculation.
In the effective lagrangian approach, the one-loop contributions to
gauge coupling constants combine one-loop diagrams from the tree-level
lagrangian (\ref{treelag}), which involve massless loops only, and
tree diagrams from the one-loop corrections to the effective
lagrangian ${\cal L}_W$, which are local corrections to (\ref{treelag}).
Consider for instance the modulus-dependent corrections to
gauge couplings at one-loop [\ref{DFKZ1}]. Using
the tree-level effective lagrangian,
one can construct a triangle diagram with two external gauge bosons and
one composite K\"ahler connection $K$. This `K\"ahler anomaly' can
be represented by a non-local superfield contribution
to the effective action
$S_\Gamma$:
\beq
\label{nonlocal}
{\cal L}_{nl}^{E_8} = - \frac{A}{4}
\left[{1 \over 3}
W^A_{E_8} W^A_{E_8} {\cal{P}}_C
  {K}\right]_F \,,
\eeq
where
\beq
\label{anomaly}
A\equiv {3C(E_8) \over {8 \pi^2}}.
\eeq
$C(E_8)$ is the quadratic Casimir of $E_8$
and ${\cal{P}}_C$ is the non-local projector of a vector multiplet
into a chiral one. For a vector multiplet $H$, ${\cal P}_C H$ is chiral,
while ${\cal P}_C\varphi=\varphi$,
${\cal P}_C\ov\varphi=0$ if $\varphi$ is chiral.
In global supersymmetry, this projector is
${\cal{P}}_C = -(16\Box)^{-1}\ov{\cal D}^2
{\cal D}^2$.
An equation similar to (\ref{nonlocal})
can be written for $E_6$ gauge fields but we prefer to
write explicitly only the hidden $E_8$ part because it
is the one we will use later. Notice that under K\"ahler
transformations
(\ref{ktransf}), the non-local term (\ref{nonlocal}) is not invariant but
generates an anomaly.  Its variation is local and contains
the terms:
\beq
\label{anomalies}
{A \over 3}
\left[
{1\over8}( \varphi + \ov{\varphi} ) F^A_{\mu\nu} F^{A\,\mu\nu}
+ {i\over8}
( \varphi - \ov{\varphi} ) F^A_{\mu\nu} \tilde{F}^{A\,\mu\nu}
\right].
\eeq
The string calculations [\ref{DKL}, \ref{ANT}]
show that in orbifolds without threshold
corrections,
this ano\-ma\-ly is cancelled by the local Green-Schwarz term
[\ref{DFKZ1}, \ref{CO}, \ref{CO2}]
\beq
\label{gsterm}
{\cal L}_{GS}
=
\frac{A}{4}{1\over3}\left[\hat{L}_{E_8} {K} \right]_D \,,
\qquad\qquad \hat{L}_{E_8} = L - 2\Omega_{E_8},
\eeq
which is a one-loop correction to ${\cal L}_W$. Actually, in theories
without threshold corrections, the Green-Schwarz lagrangian
(\ref{gsterm}) is determined from the one-loop anomaly (\ref{nonlocal})
and the information that the anomaly has to be cancelled in strings.

In the effective action,
gauge kinetic terms will receive local contributions from
the sum of (\ref{nonlocal}) and (\ref{gsterm}):
\beq
\label{F2oneloop}
\begin{array}{rcl}
{\cal L}_{nl} &\longrightarrow& -{1\over4} \left[
-{1\over6}A K\right] F^A_{\mu\nu} F^{A\,\mu\nu},
\crbig
{\cal L}_{GS} &\longrightarrow& -{1\over4} \left[
+{1\over6}A K\right] F^A_{\mu\nu} F^{A\,\mu\nu},
\end{array}
\eeq
dropping the index $E_8$. These two contributions cancel and gauge
kinetic terms are $K$--independent in orbifolds without threshold corrections,
as they should be.

Since our discussion is based on conformal supergravity,
we have to also worry about anomalies of
conformal transformations, which are expected to be related to
the renormalization-group behaviour of physical quantities.
At one-loop, the variation of the running $E_8$ gauge coupling constant
under a change $M \longrightarrow \lambda M$ of the scale is
\beq
\label{lambda}
\delta (g^{-2}) = A \log\lambda.
\eeq
In an effective lagrangian, this means that
$$
\delta \left( -{1\over4}{1\over g^2} F_{\mu\nu}^A F^{A\,\mu\nu} \right)
= -{1\over4}A \log \lambda F_{\mu\nu}^A F^{A\,\mu\nu}.
$$
This behaviour under scale transformations is precisely obtained
in the component expansion of the anomalous non-local expression
\beq
\label{nonlocalp1}
-{A\over4} \left[ W^A_{E_8} W^A_{E_8} {\cal P}_C \log \hat L \right]_F \,,
\eeq
which includes the gauge kinetic term
$$
{1\over4} {A\over2}\, \log C \, F_{\mu\nu}^A F^{A\,\mu\nu}.
$$
Notice that the expression (\ref{nonlocalp1}) is K\"ahler invariant,
in contrast with the other possibility
$$
-{A\over4} [W^A_{E_8} W^A_{E_8} \log S_0]_F \,,
$$
which would bring unwanted contributions to K\"ahler anomalies.
This argument suggests, as explained in [\ref{DFKZ2}], that by analogy
with the treatment of K\"ahler anomalies, one should complete
(\ref{nonlocal}) with
\beq
\label{nonlocalp}
\Delta_{\rm sc}{\cal L}_{nl}^{E_8} =
-{A \over 4} \left[
W^A_{E_8} W^A_{E_8} {\cal{P}}_C
 \log\frac{\hat L}{\mu^2}\right]_F  \,,
\eeq
to take conformal anomalies into account. The parameter $\mu$ can be
viewed either as a
mass scale manifesting the breaking by the anomaly of conformal
invariance, in which case it can be identified
with the renormalization group
running scale, or as a scale factor similar to $\lambda$ in eq.
(\ref{lambda}).

Again, by analogy with K\"ahler symmetry,
the next step is to add to the effective lagrangian
a local Green-Schwarz term able to cancel
these anomalies. As in [\ref{DFKZ2}]
and for reasons which will become clear in the next subsection, we
will use
\beq
\label{gstermp}
\Delta_{\rm sc}{\cal L}_{GS}
=
{A\over4}\left[\hat{L}\ {\log\frac{\hat L}{\mu^2}} \right]_D \,.
\eeq
Therefore, to appropriately use the superconformal approach for the
Wilson action we should add ${\cal L}_{GS}+\Delta_{\rm sc}{\cal L}_{GS}$ to
the tree-level effective lagrangian (\ref{treelag}),
providing an
extra correction to the gauge coupling $g_{W}^{-2}$.
Again $g_{\G}^{-2}$ does not change (at one loop),
as we will see, which is consistent with
the full string calculation. Therefore we can say that at the
scale $\mu$ the Wilson
lagrangian is
\beq
\label{waction}
{\cal L}_W=\L_0+\L_{GS}+\Delta_{\rm sc}\L_{GS},
\eeq
which is the one
corresponding to the full string calculation,
whereas the effective
lagrangian is
\beq
\label{effaction}
{\cal L}_\Gamma= {\cal L}_W
+\L_{nl}+\Delta_{\rm sc}\L_{nl} .
\eeq

To summarize the results of this subsection,
at one-loop the Wilson lagrangian is
\beq
\label{wilson}
\L_W=-\frac{1}{\sqrt 2}
\left[(S_0 \ov{S}_0)^{3/2} {\hat{L}}^{-1/2}
      e^{-K/2}\right]_D+
\left[S_0^3 w \right]_F
+{A \over 4}\left[\hat{L}\ {\log\frac{e^{K/3}
{\hat L}}{\mu^2}} \right]_D \,,
\eeq
and the effective lagrangian reads
\beq
\label{effective}
\L_\Gamma=\L_W - {A \over4}
\left[ W^A_{E_8} W^A_{E_8} {\cal{P}}_C
 \log\frac{e^{K/3}\hat{L}}{\mu^2}\right]_F \,.
\eeq
Notice that $\L_W$ is neither conformal nor K\"ahler
invariant as it should be.
In particular the superpotential cannot be absorbed
into the K\"ahler potential.
The introduction of the scale $\mu$ is of course not
compatible with superconformal
symmetry. A constant like $\mu$ should have zero conformal weight
to be a supermultiplet. But a change
in the value of $\mu$ can be
compensated by a variation of the quantity
$z_0\ov{z}_0$, which appears like a `reference scale'. The
statement that physical quantities in $\SG$ are
$\mu$--independent can then be translated into an analogous statement
on the behaviour of physical quantities under a change of the compensator,
which can be expressed in terms of superconformal invariant
expressions, up to anomalies.

\subsection{Renormalization Group and Invariant Scale}

We turn now to a discussion of the renormalization-group
equations that can be derived from the results of the
previous subsection and compare with the standard
expressions in supersymmetric field theories.

For a super-Yang-Mills theory with gauge group $G$ and without matter,
the exact renormalization-group equation is known [\ref{J}]:
\beq
\label{betaexact}
\mu{d\over d\mu} g^{-2} = {3C(G)\over 8\pi^2}
{1\over 1-{C(G)\over8\pi^2}g^2}.
\eeq
The renormalization of the gauge coupling constant
between the scales $\mu$ and $M$ is then
given by:
\beq
\label{ggamma}
\frac{1}{g_{\G}^2(\mu)}= \frac{1}{g_{\G}^2(M)}
+\frac{3C(G)}{16\pi^2}\log
\frac{\mu^2}{M^2}+\frac{C(G)}{8 \pi^2}\log
\frac{g_{\G}^2(M)}{g_{\G}^2(\mu)}+\Delta ,
\eeq
where $C(G)$ is the
quadratic Casimir of the group $G$ and
$\Delta$ represents the possible threshold corrections due to
the contributions of heavy fields to the loop
calculations. We only consider here models for which these threshold
corrections vanish.
We will identify $\mu$ with a physical scale defining for
instance the normalization of some three-point amplitude,
and $M$ with the ultraviolet cut-off of the effective theory
procedure.
In (\ref{ggamma}), we have used the
index $\Gamma$ to indicate
that  $g_\G$ is the physical gauge coupling
constant which appears
in gauge kinetic terms of the effective action $S_\G$
at the scale $\mu$.
It gets corrections to all loops in perturbation theory.

The gauge
coupling $g_W$ in the Wilson action $S_W$ is the bare coupling at the
scale $M$ and all the quantum corrections to it
give $g_\G$. In pure supersymmetric Yang--Mills
theory [\ref{SV}],
it has been found that $g_W$ {\it does not}
get corrections beyond one loop, {\it i.e.} the renormalization group
equation
reads simply
\beq
\label{rgew}
\frac{1}{g_{W}^2(\mu)}=
\frac{1}{g_{W}^2(M)}
+\frac{3C(G)}{16\pi^2}\log
\frac{\mu^2}{M^2},
\eeq
unlike $g_\G$ [see (\ref{ggamma})].
At a given scale $\mu$, the relation between $g_W$ and $g_\G$ is
\beq
\label{sv1}
\frac{1}{g_{W}^2(\mu)}= \frac{1}{g_{\G}^2(\mu)}
-\frac{C(G)}{8\pi^2}\log\frac{1}{g_{\G}^2(\mu)} ,
\eeq
making equations (\ref{rgew}) and (\ref{ggamma}) equivalent to all
loops [\ref{SV}].

We will now discuss these issues in the
string theory case. The natural cut-off scale $M^2$ is in our case
$z_0\ov{z}_0$.
{}From equation (\ref{gauge}) we can see that the Wilson gauge
coupling constant corresponding
to lagrangian (\ref{wilson}) is
\footnote{The coupling in (\ref{gawilson}) from the Wilson action
${\displaystyle \frac{1}{g_W^2(z_0\ov{z}_0)} }$ coincides with
the coupling
${\displaystyle 2(s+\ov{s}) }$ one obtains in the dual
formalism.}
\beq
\label{gawilsonold}
\frac{1}{g_W^2(z_0\ov{z}_0)}=
U-\frac{A}{3}\log{U}+\frac{A}{2}\log{\frac{z_0\ov{z}_0}{\mu^2}}
+{A\over2}\left( 1-{1\over3}\log 2\right)
,
\eeq
where $U$ is given by (\ref{gbare}), and the running term on the
right-hand side of eq. (\ref{gawilson}) is provided by the
superconformal anomaly cancelling term (\ref{gstermp}).
The last constant can be absorbed in a redefinition of the
parameter $\mu$. If we rescale $\mu \longrightarrow e^a\mu$ in lagrangian
(\ref{wilson}), the
anomaly term generates a contribution
$$
-{1\over2}aA[\hat L]_D = aA[\Omega]_D + {\rm total\,\, derivative},
$$
which is proportional to a pure super-Yang-Mills lagrangian. In particular,
it contains a term
$$
{1\over4} aA F_{\mu\nu}^A F^{A\,\mu\nu},
$$
equivalent to a change $g_W^{-2} \longrightarrow g_W^{-2} - aA$
of the gauge coupling constant.
Choosing $a={1\over2}(1-{1\over3}\log 2)$ allows
to replace (\ref{gawilsonold})
by
\beq
\label{gawilson}
\frac{1}{g_W^2(z_0\ov{z}_0)}=
U-\frac{A}{3}\log{U}+\frac{A}{2}\log{\frac{z_0\ov{z}_0}{\mu^2}}
, \qquad\qquad A= {3C(E_8)\over 8\pi^2},
\eeq
which is the expression we will use.

Formally, this equation is equivalent to a renormalization group equation
(RGE) for the running of the Wilson coupling constant
from $z_0\ov{z}_0$ to $\mu^2$. A comparison with the field theory result
(\ref{rgew}) suggests the identification
\beq
\label{gwilson}
\frac{1}{g_W^2(\mu)}=U-\frac{A}{3}\log{U} \,,
\eeq
which is not a physical quantity but only a bare parameter of
${\cal L}_W$.

Now, since (\ref{effective}) is the effective lagrangian, the coefficient
of gauge kinetic terms in ${\cal L}_\Gamma$
will provide the physical effective coupling constant at the scale
parameter $\mu$, $g_\Gamma(\mu)$.
A straightforward calculation shows that
\beq
\label{geff}
\frac{1}{g_\G^2(\mu)}=U \,.
\eeq
This important result indicates that the expectation
value of the quantity $U$, which is
a function of scalar fields of well-defined string origin like $C$
or $T$, is the {\it physical, loop-corrected} gauge coupling
constant for the $E_8$ sector of the gauge group.

Clearly, the eqs. (\ref{gwilson}) and (\ref{geff})
coincide with relation (\ref{sv1}),
\beq
\label{sv}
\frac{1}{g_W^2(\mu)}=
\frac{1}{g_\G^2(\mu)}-\frac{A}{3}\log\frac{1}{g_\G^2(\mu)}\,,
\eeq
which gives the all-order RGE for $g_\G^{-2}$ and has been
derived in the context of supersymmetric field theories [\ref{SV}].

This justifies
the result of [\ref{DFKZ2}] where all-loop RGE's were obtained from
a one-loop calculation and using the GS terms
above together with duality with the chiral multiplet
formalism. As in the super-Yang-Mills case, the Wilson
gauge coupling satisfies the
one-loop RGE (\ref{gawilson}).

Another way to derive the Wilson lagrangian (\ref{wilson}) is to
start with the tree-level gauge coupling $U$, given in eq.
(\ref{gbare}) and obtained from lagrangian
(\ref{treelag}) using eq. (\ref{gauge}), and then impose
the all-order relation (\ref{gawilson}). By integrating (\ref{gauge}),
one then recovers the loop-corrected lagrangian (\ref{wilson}).

We wish to stress that if $g_W^{-2}$
does not get corrected to higher loops, as it happens
in pure super-Yang-Mills, the Wilson lagrangian above
would be valid to all loops in string theory. There is in fact
a non-renormalization theorem limiting the moduli dependence
of the gauge coupling beyond one-loop [\ref{ANT}] in the dual
formulation in terms of chiral fields. This can give concrete
constraints to the higher loop corrections to the Wilson
action.

The effective lagrangian (\ref{effective}) provides
the $E_8$ renormalization from the cut-off scale $z_0\ov{z}_0$ to the
{\em low} scale $\mu$. Notice that in the range
$z_0\ov{z}_0 \longrightarrow \mu^2$
the renormalization of the gauge couplings is triggered by {\it stringy}
effects encoded in (\ref{wilson}) and (\ref{effective}).
This running is generated by the (stringy) anomaly cancellation
mechanism of (\ref{effective}). Since this mechanism is universal,
it can only take care of a single
gauge coupling evolution. In the case of the
$Z_3$ and $Z_7$ orbifolds we are considering, it describes
the evolution of the {\it hidden} $E_8$, which will be useful for the
gaugino condensate mechanism of supersymmetry breaking.

The equation (\ref{geff}) indicates that the vacuum expectation
value of the field-dependent quantity $U$ is the physical gauge
coupling constant in the effective action
computed with a string infra-red cut-off $\mu$.
{}From its definition (\ref{gbare}), $U$ depends
on $C$, $K$ and on the compensator $z_0$. In the Poincar\'e
theory, the compensator itself is a function of $C$, $K$ and
$\kappa^2$,
so that the physical gauge coupling constant in the effective
lagrangian is a function of the expectation values of $C$ and $K$.
In the dual theory, with the chiral multiplet $S$ instead of $L$,
the general (all-order) result (\ref{fis4S}) indicates that
the expectation value of $\Re s$ is the bare coupling constant,
a non-physical quantity. The chiral multiplet $S$ appears to be
an artifact of the effective field theory, while the linear multiplet
$L$ is directly related to the physical fields of the string
theory. Even if the two theories are formally equivalent by duality,
loop corrections introduce clear conceptual differences
in their interpretation.

The Einstein term in the effective lagrangian can be computed using
eq. (\ref{Einstein1}). The non-local parts do not contribute and one
obtains
$$
-{1\over2} [U+{A\over6}] C eR.
$$
Fixing conformal symmetry by the requirement of canonical Einstein
terms leads to the relation
\beq
\label{Acompfix}
U = {1\over\kappa^2 C} - {C(E_8)\over 16\pi^2} =
{1\over g^2_\Gamma(\mu)},
\eeq
which indicates that the physical effective gauge coupling constant in
the effective, loop-corrected lagrangian is controlled by the
expectation value of the scalar field $C$. Loop corrections only
introduce a constant shift [compare with (\ref{g2Poin})].

By construction [see eq. (\ref{gawilson})], $U$ satisfies the RGE
$$
\mu{d\over d\mu} U = {A\over 1-{A\over 3U}}\,,
$$
which is identical to eq. (\ref{betaexact}). The reaction of $U$
to a change in the arbitrary value of the string cut-off
$\mu$ is in agreement with all-order RGE, which implies that
physical quantities (scattering probabilities)
computed in the effective lagrangian
are independent of $\mu$.

It is now easy to write down the renormalization-group
invariant scale $\Lambda_{E_8}$ as a
function of the fields of the theory. It can be constructed
either using the Wilson gauge coupling (\ref{gawilson}) or the effective
gauge coupling (\ref{geff}), as
\beq
\label{inv}
{\displaystyle
\Lambda_{E_8}^3 =(z_0\ov{z}_0)^{3/2}
e^{-\frac{3}{A}\frac{1}{g_W^2(z_0\ov{z}_0)}}
=\mu^3 U e^{-\frac{3}{A}U}. }
\eeq
The scale $\Lambda_{E_8}$ is independent of the choice of $\mu$:
using the exact RGE (\ref{betaexact}),
one easily checks that $\mu{d\over d\mu}\Lambda^3_{E_8}=0$.
The invariant parameter $\Lambda_{E_8}$ characterizes the strength
of $E_8$ gauge interactions. It is a physical quantity, independent of
the choice of cut-off $\mu$.
Moreover, the real, field-dependent quantity $\Lambda_{E_8}^3 /
\mu^3$ is the lowest component of a real vector supermultiplet.
This is due to the fact that $U$, as defined by eq. (\ref{gbare}),
is itself the lowest component of the real, K\"ahler and conformal
invariant, vector supermultiplet
$$
2\left( {S_0\ov S_0 \over 2\hat L e^{K/3}}\right)^{3/2},
\qquad
K = K(\Sigma, \ov \Sigma e^V).
$$

With the compensator fixing condition (\ref{Acompfix}), the
renormalization-group invariant scale becomes
\beq
\label{inv2}
\Lambda_{E_8}^3 =
\mu^3 e^{1\over2} \left( {1\over\kappa^2C}- {C(E_8)\over 16\pi^2}\right)
{\rm exp}\left( -{8\pi^2 \over C(E_8)} {1\over \kappa^2 C}\right),
\eeq
a formula of direct interest in the discussion of gaugino condensation.

At the string-tree level, one can define a
compactification scale by the expression
\beq
\label{mcomp}
M^2 = (2C)e^{K/3},
\eeq
which is the translation of the more usual
relation $M^2 = {1\over\kappa^2}[(S+\ov{S})(T+\ov{T})]^{-1} =
{1\over\kappa^2}(S+\ov S)^{-1}e^{K/3}$,
which holds in the
dual formalism [\ref{DRSW}].

It is consistent to use expression (\ref{mcomp}) inside the logarithmic
term of (\ref{gawilson}) since we are performing a one-loop calculation.
Using (\ref{mcomp}) in (\ref{gawilson}), we obtain
\beq
\label{unification}
\frac{1}{g_W^2(z_0\ov{z}_0)}=\frac{1}{g_\G^2(M)} \,.
\eeq
Equation (\ref{unification}) shows that while the effective gauge
couplings unify (modulo threshold corrections) at the compactification
scale, the Wilson gauge couplings unify at the Planck scale ({\it i.e.}
$z_0\ov{z}_0$) with the {\it same} value. Finally eq. (\ref{unification})
shows that the effect of gauge coupling running in (\ref{gawilson}) can
be absorbed in the very definition of $M$.

\nsect{String Component Actions and Gaugino \hfill\break
Condensation}

In the previous section, we have constructed the loop-corrected
effective actions
for particular string compactifications in the supermultiplet
language. In order to obtain further
information from these actions, we will present here their component
expressions using the general formalism developed in sections 3
and 4. As mentioned previously,
we will
concentrate on the couplings which are relevant to  the
study of supersymmetry breaking and then restrict to the scalar
kinetic terms and potential, including gaugino bilinear contributions.
We will first present these couplings for the loop-corrected
Wilson action which is neither conformal
nor K\"ahler invariant.
The effective potential will be obtained by summing
the contributions from the Wilson lagrangian and from
the non-local actions (\ref{nonlocal}) and (\ref{nonlocalp}),
which provide local gaugino-dependent contributions to
the effective potential, restoring the invariance under both K\"ahler
and conformal transformations, before compensator fixing.
In other words, the effective potential is extracted from the
effective lagrangian (\ref{effective}).

We would like to emphasize that there is a technical
difficulty
with the linear multiplet formulation as treated in the
superconformal approach. Contrary to the purely chiral multiplet
action (\ref{Cremmer1}) in which
the auxiliary field equations can be solved and the compensator
can be fixed to provide a general form of the scalar potential as a
function of $\cal G$  and $f_{AB}$,
this cannot be done for the linear multiplet action. The reason
is that since the linear multiplet has conformal weight two,
the function $\Phi$ itself
depends on the compensator. The solution to
equation (\ref{Lcomp}), which fixes the compensator,
is then only implicit. Also,
the equations of motion for the auxiliary fields have to be solved
for each $\Phi$ to provide the scalar potential. Furthermore,
the Wilson action is not conformal invariant
and the component expressions (\ref{LKIN}) and (\ref{LAUX})
cannot
be used directly for the Green-Schwarz terms which are anomalous.
These Green-Schwarz contributions can however easily be computed
using the truncated supermultiplets (\ref{embed}).
The complete expressions are given in the appendix, together
with the truncated component expansions of the non-local
terms appearing in the loop-corrected effective action
(\ref{effective}).

It should be stressed that our component expressions are based upon
lagrangians (\ref{wilson}) and (\ref{effective}),
which include all-order corrections in $\hat L$ but
not in chiral matter.

\subsection{String Effective Actions in Components}

It is straigthforward to derive the tree-level string action
in components knowing the general results from the previous
sections. The tree-level lagrangian (\ref{treelag})
is a particular case of the class (\ref{useless1}) for which
$F(u)=-\frac{1}{\sqrt 2}u^{-3/2}$. We can easily solve the compensator
fixing
condition (\ref{Lcomp}) giving:
\beq
\comp=(2\kappa^{-4}Ce^K)^{1/3} .
\eeq
With eq. (\ref{POTD}), the scalar potential
including gaugino bilinears is:
\beq
\label{thispot}
\kappa^{2}V=2C e^K\left( K_{z\ov z}^{-1}\left\vert
w_z+wK_z\right\vert ^2-3\left\vert w\right\vert ^2\right)
+\left\vert(2Ce^K)^{1/2}w+\frac{1}{2C}(\ov\lambda_R\lambda_L)
\right\vert ^2 .
\eeq
This exhibits the known property that the scalar potential
is positive definite for no--scale K\"ahler
potentials for which the first term cancels.
As discussed in [\ref{DRSW}],
gaugino condensation can break supersymmetry with vanishing
cosmological
constant provided there is a non-zero vacuum expectation value
for the superpotential. Notice that the potential (\ref{thispot})
is equivalent to the
more familiar expression in terms of the dual $S$ field as can
be seen
by using the duality transformation which in this case amounts
to set
$(\kappa^2 C)^{-1}=2(s+\ov s)$
[see eq. (\ref{g2Poin})]. It can also be verified that under
 this
substitution, the scalar kinetic terms obtained using eq.
(\ref{LKIN}) take the same form as in
[\ref{W}].

We will next consider the loop-corrected action (\ref{wilson}), which
we  write as:
\beq
\label{wilsond}
\L_W=\L_0+ \frac{A}{4}\left[\hat{L}\ {\log\frac{e^{K/3}{\hat L}}{\mu^2}}
\right]_D .
\eeq
where $\L_0$ is the tree-level lagrangian (\ref{treelag}) which is
conformal and K\"ahler invariant.
The loop term can be written as
\beq
\label{wilsont}
\begin{array}{rcl}
{A\over4}\left[\hat{L}\ {\log e^{K/3}\frac{{\hat L}}{\mu^2}} \right]_D&=&
\frac{A}{12}\left[S_0 \ov{S}_0  \frac{\hat{L}}{S_0 \ov{S}_0 }\ {K}
\right]_D+ {A\over4}\left[\hat{L}\ {\log\frac{\hat L}{\mu^2}}\right]_D
\crbig
&=& {A\over4}\left[S_0 \ov{S}_0  \frac{\hat{L}}{S_0 \ov{S}_0 }\log
\left(e^{K/3}
\frac{\hat{L}}{S_0 \ov{S}_0 }\right)\right]_D+
{A\over4} \left[\hat L\log
(\frac{{S_0 \ov{S}_0 }}{\mu^2})\right]_D.
\end{array}
\eeq
In both expressions, the first term is conformal invariant  and
the second is anomalous (only the first term in the second expression
 is both K\"ahler and
conformal invariant and has the form (\ref{useless2}) with
 $F(u)= {A\over4} \log u$).
The component expressions for both anomaly terms are
given in the appendix. The auxiliary field part of the
lagrangian can always be written in the form
(\ref{LAUXD}), the solutions for $f$, $f_0$ and $V_{aux}$
 are as in
(\ref{SOL}) and (\ref{POT}), but with different expressions
for the coefficients in (\ref{ABC}).

For definiteness, let us use the full action (\ref{wilsond}) with
the one-loop term
written as in the second expression (\ref{wilsont}). We see
that
the conformal and K\"ahler invariant part has the form of
(\ref{useless2})
with
\beq
F(u)=-\frac{1}{\sqrt 2}u^{-3/2}+ {A\over4} \log u ,
\eeq
which we can use in
(\ref{ABC}). The contribution from the anomaly term
has the effect of changing some coefficients in (\ref{ABC}):
\beq
\begin{array}{rcl}
{\c D}&\longrightarrow &{\c D}- {A\over4}, \crbig
{\c E}&\longrightarrow &{\c E}- {A\over4} z_0^{-1}
(\ov\lambda_R\lambda_L),
\end{array}
\eeq
[see eq. (\ref{A2}) in the appendix].
The Wilson scalar potential with the loop corrections and
before fixing the compensator reads then
\beq
\label{potcomp}
\begin{array}{rcl}
V_W&=&\comp^3  \left(\frac{1}{6}A C +UC\right)^{-1} K_{z\ov z}^{-1}
\left\vert w_z+w K_z-\frac{A}{12}z_0^{-3} K_z (\ov\lambda_R\lambda_L)
\right\vert ^2\crbig
&&
-2(UC)^{-1}\comp^3\left\vert w - \frac{1}{4}
Uz_0^{-3}(\ov\lambda_R\lambda_L)\right\vert ^2
-\frac{3}{8C}(\frac{1}{3}A -U)\left\vert (\ov\lambda_R\lambda_L)
\right\vert ^2 ,
\end{array}
\eeq
where $U = 2\left({z_0\ov z_0 \over 2Ce^{K/3}}\right)^{3/2}$ is
the quantity already introduced in (\ref{gbare}) whose expectation
value is the physical coupling $g^{-2}_\Gamma$.

The requirement that the Einstein term, which has already been
evaluated in the previous subsection, is canonically normalized
gives the compensator fixing condition (\ref{Acompfix}),
$C(U +{A\over6}) = \kappa^{-2}$, or
\beq
\label{Acompfix2}
z_0 \ov z_0 =(2Ce^K)^{1/3} \left(\frac{1}{\kappa^2}-\frac{A}{6} C\right)
^{2/3}.
\eeq
Therefore, the scalar potential after fixing the compensator takes the
form:
\beq
\label{POTt}
\begin{array}{rcl}
\kappa^2 V_W&=&2Ce^K(1-\frac{\kappa^2}{6}A C)^{2}
K_{z\ov z}^{-1} \cdot \crbig
&& \cdot
\left\vert w_z+wK_z-\frac{A}{12}(2Ce^K)^{-1/2}(\frac{1}{\kappa^2}-
\frac{1}{6}A C)^{-1}
(\ov\lambda_R\lambda_L)K_z\right\vert ^2\crbig
&&
-{4}Ce^K(1-\frac{\kappa^2}{6}A C)\left
\vert w-\frac{1}{2}(2C)^{-3/2}e^{-K/2}(\ov\lambda_R\lambda_L)
\right\vert ^2\crbig
&&+\frac{3}{8C^2}
(1 -{A\over2}\kappa^2 C)\left\vert (\ov\lambda_R\lambda_L)
\right\vert ^2.
\end{array}
\eeq
Notice that by  lack of K\"ahler invariance, the superpotential
and K\"ahler potential {\it cannot} be combined
 into a single K\"ahler-invariant function. It is also interesting to
observe that
there is no $\mu$ dependence in $V_W$, and
the same will hold for the scalar
kinetic terms. Actually, the only $\mu$ dependence of the Wilson
action is in the gauge kinetic terms, as we saw in the previous
chapter.

The scalar kinetic terms can be obtained by shifting  ${\c D}$ as
indicated above. The
final form after fixing the compensator is \footnote{
Notice that the kinetic lagrangian, eq. (\ref{effectivekin}),
unlike the Wilson potential, eq. (\ref{POTt}), is invariant under K\"ahler
transformations, thanks to conditions (\ref{Lcomp2}).
}
\beq
\label{effectivekin}
\begin{array}{rcl}
e^{-1}{\cal L}_{kin}&=& -{1\over\kappa^2}
K_{z\ov z}(\partial_\mu z)(\partial^\mu \ov z)
-\frac{1-{1\over2}\kappa^2 A C}{4\kappa^2 C^2}\left(
{1-\frac{\kappa^2}{6}AC}\right)^{-1}(\partial_\mu C)(\partial^\mu C)
\crbig
&&
+\frac{1}{4\kappa^2 C^2}
\left(1-\frac{2\kappa^2}{3}AC\right)v_\mu v^\mu
-\frac{i }{12}Av^\mu\left(K_z\partial_\mu z - {\rm h.c.}\right).
\end{array}
\eeq
The kinetic terms for
the chiral scalar fields are still given by the tree-level
K\"ahler potential. The positivity of these kinetic
terms for $C$
sets the allowed range for that field. The
novel feature is that we obtain an `off-diagonal'
term mixing
the chiral fields and the antisymmetric tensor represented
by $v_\mu$. Similar mixing terms were obtained
in [\ref{GT}, \ref{ABGG}]. There are however no mixed terms with
$C$.

Finally, to obtain the effective potential,
we have to consider the local contributions of the  non-local
lagrangian. The non-local terms only contribute by
gaugino-dependent terms given by
\beq
-\frac{A}{4C}\left\vert (\ov\lambda_R\lambda_L)
\right\vert^2
           +\frac{A}{12}\left[ K_z f (\ov\lambda_R\lambda_L)+{\rm h.c.}
\right]
\eeq
[see the appendix, eq. (\ref{A3})], and so they do not
contribute to the kinetic lagrangian (\ref{effectivekin}).
We can see that this has the net effect of shifting
\beq
\begin{array}{rcl}
{\c F}&\longrightarrow &{\c F}+\frac{A}{12} K_z f
(\ov\lambda_R\lambda_L),
 \crbig
{\c L}_{4\lambda}&\longrightarrow &{\c L}_{4\lambda}-
\frac{A}{4C}\left\vert (\ov\lambda_R\lambda_L)\right\vert ^2 ,
\end{array}
\eeq
in the expressions (\ref{ABC}).
The effective potential before compensator fixing becomes
\beq
\label{POTeff}
\begin{array}{rcl}
V_{eff}&=&\comp^3  \left(\frac{1}{6}A C +UC \right)^{-1}
K_{z\ov z}^{-1}
\left\vert w_z+w K_z\right\vert ^2 -
2\comp^3\left\vert w\right\vert ^2(CU)^{-1}\crbig
&&+\frac{1}{2 C}\left[(\ov\lambda_R\lambda_L)z_0^3 w+ {\rm h.c.}
\right]
 + \frac{1}{4C}\left({A\over2}+U \right)
\left\vert (\ov\lambda_R\lambda_L)\right\vert ^2 .
\end{array}
\eeq
And fixing the compensator using (\ref{Acompfix2})
produces the effective potential
\beq
\label{veffd}
\begin{array}{rcl}
\kappa^2 V_{eff}&=&2Ce^K\left(
1-\frac{\kappa^2}{6}AC\right)^2\left( K_{z\ov z}^{-1}\left\vert
 w_z+
K_z w\right\vert ^2- 3\left\vert w\right\vert ^2\right)\crbig
&&+\left(
1-\frac{\kappa^2}{6}AC\right)(1-{1\over2}\kappa^2 AC)
\left\vert (2Ce^K)^{1/2}
 w+\frac{1}{2C}(1-{1\over2}\kappa^2 AC)^{-1}
(\ov\lambda_R\lambda_L)\right\vert ^2
\crbig
&&-\frac{A^2\kappa^4}{24}
(1-{1\over2}\kappa^2 AC)^{-1}\left\vert (\ov\lambda_R\lambda_L)
\right\vert
^2 ,
\end{array}
\eeq
in the loop-corrected Poincar\'e theory.


\subsection{Gaugino condensation}

In this section we will discuss the issue of supersymmetry
breaking by gaugino condensation in the context of string
effective actions with a linear multiplet. This phenomenon is
expected to happen [\ref{AKMRV}] when the gauge
coupling of the hidden gauge
group becomes strong [\ref{DIN1},\ref{DRSW}].
Being a non-perturbative effect, its
dynamics is not completely understood.

All previous discussions on the subject have been done using the
dual formalism with only chiral multiplets. There are
at least three different methods
which have been used to incorporate
the effects of the gaugino condensate
in the effective theory. One approach is the simple
substitution of the condensate as an expectation value
of gaugino bilinear operators in the component
action [\ref{FGN}], $\ov{\lambda}_L \lambda_R \sim \Lambda^3$,
where $\Lambda$ is the scale of condensation, which in string
theory is a field-dependent quantity. Another way is replacing
the condensate
in the superfield action, generating an
effective superpotential for the dilaton field
determined by symmetry arguments [\ref{DRSW}],
$w(S)\sim exp(-6S\beta)$, where $\beta$ is the coefficient of
the beta function of the hidden gauge group.

A third approach is the
formulation of an effective supersymmetric theory, below
the scale of condensation, where the gauge invariant condensate
$W^AW^A$ is described in terms of
a new dynamical superfield $Z$ determined
upon minimization of the scalar potential [\ref{VY}]
(see also [\ref{MR}]).

Let us briefly discuss these approaches in the linear multiplet
formulation. We can also carry out the naive
substitution of the condensate in the component action.
Notice that we have the same situation as
in the global supersymmetry case,
discussed in section 2, in which a non-vanishing gaugino
condensate amounts to a non-supersymmetric
shift in the `$\theta\theta$ component' of the
Chern-Simons multiplet.

In the absence of a superpotential, the
scalar potential (\ref{veffd}) is
\beq
\label{veffnow}
\kappa^2 V_{eff}=(2C)^{-2}(1+{1\over3}\kappa^2 AC)\mid
\ov{\lambda}_L \lambda_R \mid^2 ,
\eeq
where the factor in front of $\mid \ov{\lambda}_L
\lambda_R\mid^2 $ is positive definite due to the positive
kinetic energy conditions. We now replace gaugino bilinears by
expectation values,
\beq
\ov{\lambda}_L \lambda_R \sim \Lambda^3_{E_8},
\eeq
where $\Lambda_{E_8}$ is the renormalization group and K\"ahler
invariant scale defined in (\ref{inv}).
After fixing the compensator, according to eq. (\ref{inv2}),
$\Lambda_{E_8}^3 \sim \exp\{-3/A\kappa^2C\}$, and the potential for the
$C$ field has a `runaway' behaviour towards the supersymmetric
(singular) minimum $C=0$, which also corresponds to vanishing
gauge couplings. This is the
situation already encountered in the dual formalism for the case of
a single condensate \footnote{
Though the present calculation is restricted to a very
particular class of orbifold compactifications with a simple
gauge group in the hidden sector, one could expect  a
similar pattern for the potential to hold in more complicated cases,
where the hidden sector contains several gauge groups. In those
cases, for vanishing superpotential, a stable
supersymmetric minimum for $C\neq 0$ could be generated.}.

The approach of including the condensate in the superpotential
is not feasible in the linear multiplet formalism because the
superfield $L$ cannot appear in the superpotential. Even more,
since in the dual formulation the existence of a superpotential
for the $S$ field breaks the Peccei-Quinn symmetry ${\rm Im}\ S
\rightarrow {\rm Im}\ S+{\rm constant}$, the connection to the
original formulation with a linear multiplet remains unclear.
In
fact, the very existence of that symmetry is what allows to perform
the inverse duality transformation from the $S$ to the $L$ field
formalism \footnote{Notice that in the global case, discussed in chapter 2,
 a gaugino bilinear can be seen as a
linear contribution to the superpotential [eq. (\ref{neww})].
This is not the case in local supersymmetry.
As discussed in section 3 the only
superpotential for S allowed by the duality
transformation is an
{\it overall} $\exp\{-aS\}$ [see eq. (\ref{Ssuperpot})],
which is not of the form obtained
in studies of supersymmetry breaking by gaugino condensation in
the $S$ field approach.}. This raises the interesting question of
understanding which is the appropriate formalism to be used
in this case.

As for the approach proposed in ref. [\ref{VY}], we only wish
to mention here the formal similarity between the construction of
the effective action below the condensation
scale of a super-Yang-Mills theory and the r\^ole played by
the Green-Schwarz terms in string effective actions.
The Green-Schwarz counterterms are adjusted to cancel a perturbative
anomaly since strings dictate the absence of any K\"ahler
anomaly, while the effective theory constructed in [\ref{VY}]
has an anomaly behaviour dictated by the properties of dynamics at
energies higher than the condensation scale.
More details on these issues
will appear elsewhere.

\nsect{Conclusions}

In this paper, we have addressed a number of issues concerning
the coupling of one linear supermultiplet to chiral multiplets in
global and local supersymmetry. The main motivation was the
fact that this is the spectrum in four-dimensional strings. We
have applied the results to obtain general information about
the string effective actions and, also, to explicitly compute
the loop-corrected effective actions in simple orbifold
compactifications.

Let us briefly summarize the main results of this paper. First,
we have used the superconformal approach to compute the
component supersymmetric actions of one linear multiplet
coupled to chiral multiplets. It generalizes the case of only
chiral multiplets studied in ref. [\ref{CFGVP}] and reduces to
it as a particular case when the linear multiplet is decoupled.
Unlike this case, it is not possible to write a closed
expression for the lagrangian because the compensator fixing
equation cannot be solved in general. Though, we have been able
to find very general expressions before fixing the compensator
and, for particular cases (namely for string effective actions),
we have solved the compensator fixing equation and found
explicit expressions for the interesting pieces of the lagrangian.

In the linear multiplet formalism a general
loop-corrected Wilson lagrangian takes the form
\beq
\label{genlagg}
{\cal L}_W=\left[\Gamma(\hat{L},S_0,\ov{S_0},\Sigma,\ov{\Sigma}e^V)
\right]_D+\left[S_0^3 w\right]_F ,
\eeq
where $\Gamma$ is an arbitrary real function,
its functional form includes functions of the
form $\Phi(\hat L/S_0\ov{S_0},\Sigma,\ov{\Sigma}e^V)$
discussed in the text,
corrected  by Green-Schwarz counterterms as in section 5.
The gauge coupling is given by
\beq
\label{ggauge}
\frac{1}{g^2_W}=2\left [ \frac{\partial \Gamma}{\partial \hat{L}}
\right ]  _{\rm lowest \,\, component} .
\eeq
{}From (\ref{ggauge}), one can see the interesting property that it
is sufficient to obtain the loop corrections to the gauge coupling
to determine in large parts,
through integration of (\ref{ggauge}), the loop
corrections to the Wilson action. In this way, any
non-renormalization theorems on the gauge coupling [\ref{ANT}] would
translate into non-renormalization theorems on the whole
Wilson action.

Moreover, we have seen that
the linear multiplet is the supersymmetric extension of the
string loop-counting parameter. It follows rules different from chiral
or vector multiplets and is then naturally singled out by
supersymmetry. Corrections due to string loops will be reflected
in the functional dependence on $\hat L$ of the effective theory.

Concerning the superpotential in (\ref{genlagg}), it is
straightforward to see in this approach that it {\it cannot} be
renormalized to any loop in
perturbation theory, since it does
not depend on $L$ (nor $S_0$), which is the string loop-counting
parameter. This argument can be extended to any
non-perturbative effect as long as the formalism holds.
The K\"ahler potential for chiral matter
cannot either get any loop
corrections since it is also a function of only the chiral
multiplets. There is however no restriction on the couplings of chiral
matter to $\hat L$ described in the real superfield $\Gamma$,
which gets loop corrections. We saw for instance
in the case studied in section 6 that the kinetic terms
acquire a $v^\mu\partial_\mu z $ mixing which was not present
at tree-level. Nevertheless, the loop-corrected quantities
are simple functions of the tree-level K\"ahler potential $K$ also.
Then we can say that the tree-level techniques used to compute
$K$, such as the use of special geometry in $(2,2)$ models,
are still useful for obtaining the effective lagrangians
beyond tree-level.

Even though the theory is dual to one expressed only in terms of
chiral multiplets, for which the general form of the effective
action is known, the linear multiplet formalism is more
convenient to discuss the loop-corrected string effective
action, which may not have a closed form when expressed in terms
of only chiral multiplets. This point can be illustrated by
performing the duality transformation, as explained in
section 3, on the Wilson lagrangian (\ref{wilson}).
The first step of the duality
transformation is to define an equivalent lagrangian of
the form
\beq
\label{duallag}
{\cal L}={\cal L}_W(\hat{L} \rightarrow Q)-[(S+\ov{S})(Q+2\Omega)]
_D,
\eeq
where $Q$ is an arbitrary real vector superfield. By integration
of $S$ we recover ${\cal L}_W(\hat{L})$. The equation of
motion for $Q$ is
\beq
\label{rgecon}
{2}(S+\ov{S})=\tilde U-\frac{A}{3}\log \tilde U
+ {A\over2}\log\left({S_0\ov S_0\over \mu^2}\right)
+{A\over2}(1-{1\over3}\log 2),
\eeq
where now
$$
\tilde U=2 \left({S_0\ov{S}_0 \over 2Qe^{K/3}}
\right)^{3/2}.
$$
This superfield equation of motion
is formally identical to eq. (\ref{gawilsonold}),
which defines the Wilson gauge coupling constant. It is then clear
that (\ref{rgecon}) identifies the lowest component $s$ of
$S$ as
$$
2(s+\ov s) = {1\over g_W^2\comp},
$$
a non-physical (bare) parameter, while, according to (\ref{geff}),
$$
U = {1\over g_\Gamma^2(\mu)},
$$
a physical quantity.
To complete the transformation and construct the
dual lagrangian one has to solve in (\ref{rgecon}) $U$ as a
function of the chiral superfield $S$, which {\it cannot} be done
analytically.
This can only be done in a perturbative
series coresponding to the expansion of the physical
coupling $g_\Gamma$ as a function of the bare parameter $g_W$.
This argument shows that the linear multiplet formalism allows to
write all-order closed expressions in terms of physically relevant fields
and parameters, while the dual $S$-field formalism uses unphysical fields
leading necessarily to perturbative results, with the
additional difficulty of distorting the symmetry structure of
the theory.

Finally, we have discussed the issue of supersymmetry breaking
by gaugino condensation. For the case of global supersymmetry, we
found out that a non-vanishing vacuum expectation value for the
condensate breaks supersymmetry explicitly in the linear
multiplet formalism, whereas the breaking seems at
first sight spontaneous in the
dual formalism. (However the two versions are
equivalent.) For the string case, we have
pointed out the apparent
inconsistency of the breaking of the Peccei-Quinn symmetry
(necessary to generate the superpotential for the $S$ field
in the chiral approach) and
the duality transformation relating it to the linear multiplet
approach.
If the formalism with the linear multiplet extends to the
strong coupling regime, it would predict the existence of  an
exactly massless particle corresponding to the antisymmetric tensor
$b_{\mu\nu}$ which in the dual theory corresponds
to a massless `axion' field. There exists
the logical alternative that in the
strong coupling regime, it is only the $S$ field formulation which is
valid and the axion field gets a mass after supersymmetry and
the Peccei-Quinn symmetry are
broken. This is what is implicitly assumed in the literature, but at
present there is no concrete
evidence to support this assumption
and we have to take seriously the possibility
of a massless axion field
(and probably also other moduli fields) as
a consequence of an exact Peccei-Quinn symmetry
in string theory. We will present further developments on these topics in
a future publication.

\vspace{5mm}
\section*{Acknowledgements}

We would like to acknowledge useful conversations with
C. P. Burgess, S. Ferrara and L. E. Ib\'a\~nez.
One of us (MQ) wishes to thank the Institut de Physique,
Universit\'e de Neuch\^atel, for its warm hospitality while part of
this work was performed.

\newpage
\nappend{Appendix}

We give in this appendix some component expressions useful in the text.

\vskip.2truecm
{\noindent\it 1) The abelian Chern-Simons global superfield:}
\vskip.2truecm

\noindent
The component expansion of the Chern-Simons superfield (\ref{CSabel}),
for a single, abelian vector superfield $V$ and in the
Wess-Zumino gauge is:
\beq
\begin{array}{rcl}
\Omega(V) &=& -{1\over4} \big[
-\theta\theta\ov{\lambda\lambda}-\ov{\theta\theta}\lambda\lambda
-2(\theta\sigma^\mu\ov\theta)(\lambda\sigma_\mu\ov\lambda) \crbig
&& +2i\theta\theta\ov\theta\ov\lambda D
-2i\ov{\theta\theta}\theta\lambda D \crbig
&& +\theta\theta\ov\theta\ov\sigma^\mu\sigma^\nu \ov\lambda F_{\mu\nu}
+\ov{\theta\theta}\theta\sigma^\mu\ov\sigma^\nu\lambda F_{\mu\nu}
\crbig
&&
+\theta\theta\ov{\theta\theta} \left( DD
- i\lambda\sigma^\mu\partial_\mu\ov\lambda
+ i\partial_\mu\lambda\sigma^\mu\ov\lambda
-{1\over2} F^{\mu\nu} F_{\mu\nu} \right) \crbig
&& -i\theta\sigma^\mu\ov\lambda a_\mu
+i\lambda\sigma^\mu\ov\theta a_\mu
-2(\theta\sigma_\mu\ov\theta)
\epsilon^{\mu\nu\rho\sigma}a_\nu\partial_\rho a_\sigma  \crbig
&&
-{1\over2} \theta\theta\ov\theta\ov\sigma^\mu\sigma^\nu
\partial_\mu(a_\nu\ov\lambda)
-{1\over2} \ov{\theta\theta}\theta\sigma^\mu\ov\sigma^\nu
\partial_\mu(a_\nu \lambda)
\big] ,
\end{array}
\eeq
with $F_{\mu\nu}= \partial_\mu a_\nu -\partial_\nu a_\mu$.
The last two lines contain the gauge-variant terms, which include
the Chern-Simons abelian form in the $\theta\sigma^\mu\ov\theta$
component. Notice that this expression does not contain the
CP-odd expression $\epsilon^{\mu\nu\rho\sigma}F_{\mu\nu}F_{\rho\sigma}$.
Also, the lowest component without $\theta$, $\ov\theta$, vanishes,
and the highest $\theta\theta\ov{\theta\theta}$ component is the
pure super-Yang-Mills lagrangian.
This last two observations apply to the non-abelian Chern-Simons superfield
as well.

\vskip.2truecm
{\noindent\it 2) Supergravity expressions for anomaly terms:}
\vskip.2truecm

\noindent
The Green-Schwarz terms carrying the conformal anomaly
contain $D$-densities of functions of
$\hat L=L-2\Omega(V)$ and $S_0$
which do not depend on the combination
${\hat L\over S_0\ov S_0}$.
Their component expansions using truncated supermultiplets
(\ref{embed}) read
\beq
\label{A2}
\begin{array}{rcl}
[\hat L\log \hat L]_D &=&
{1\over2} C^{-1}(\partial_\mu C)(\partial^\mu C)
-{1\over2}C^{-1} v_\mu v^\mu \crbig
&& +{1\over2} C^{-1}(\ov\lambda_L\lambda_R)(\ov\lambda_R\lambda_L
) -{1\over3}CR, \crbig
[\hat L\log(S_0\ov S_0)]_D &=&
-v_\mu A^\mu
-  z_0^{-1}f_0(\ov\lambda_R\lambda_L)
-  \ov z_0^{-1} \ov f_0 (\ov\lambda_L\lambda_R),
\end{array}
\eeq
up to total derivatives. These expressions appear in the Wilson
lagrangian (\ref{wilson}), (\ref{wilsont})
and are used in the derivation
of the scalar potentials (\ref{POTt}) and (\ref{veffd}).

\vskip.2truecm
{\noindent\it 3) Non-local terms in the effective
lagrangian (\ref{effective}):}
\vskip.2truecm

\noindent
The action of the chiral projector ${\cal P}_C$ on a
truncated vector multiplet $H$
with components $(c,0,h,k,b_\mu,0,d)$ is to form a chiral truncated
multiplet ${\cal P}_C H = (z,0,f)$ with
$$
z = {1\over2}c - \Box^{-1}(d +{i\over2}\partial^\mu b_\mu
), \qquad f={1\over2} (h+ik).
$$
The component expansion of the non-local expression appearing in
the loop-corrected effective lagrangian (\ref{effective})
is then:
\beq
\label{A3}
\left[ {1\over4} W^AW^A {\cal P}_C \log \left(\hat Le^{K/3}\right)
\right]_F = {1\over4} \left[
{1\over 3} K_z f (\ov\lambda_R\lambda_L)
+{1\over 3} K_{\ov z} \ov f(\ov\lambda_L\lambda_R)
- C^{-1} (\ov\lambda_R\lambda_L)(\ov\lambda_L\lambda_R) \right],
\eeq
using the truncated multiplets (\ref{embed}).
Notice that (\ref{A3}) is local.
Also a local gauge kinetic term of the form
\beq
\label{A4}
-{1\over4} \left[ {1\over6}K + {1\over2}\log C\right] F_{\mu\nu}^A
F^{A\,\mu\nu}
\eeq
is present.

\newpage

\section*{References}
\begin{enumerate}
\item
\label{M}
J. Minahan, \NP{B298}{88}{36};
\\
W. Lerche, \NP{B308}{88}{102}
\item
\label{K}
V. S. Kaplunovsky, \NP{307}{88}{145}
\item
\label{DKL}
L. J. Dixon, V. S. Kaplunovsky and J. Louis, \NP{B355}{91}{649}
\item
\label{ANT}
I. Antoniadis, K. S. Narain and T. R. Taylor, \PL{B267}{91}{37};
\\
I. Antoniadis, E. Gava and K. S. Narain, \PL{B283}{92}{209};
\NP{B383}{92}{93}
\item
\label{linear}
S. Ferrara, J. Wess and B. Zumino, \PL{B51}{74}{239};\\
S. Ferrara and M. Villasante, \PL{B186}{86}{85}; \\
S. Cecotti, S. Ferrara and L. Girardello, \PL{B198}{87}{336};
\\
B. Ovrut, \PL{B205}{88}{455}
\item
\label{S}
W. Siegel, \PL{B85}{79}{333}
\item
\label{Betal}
P. Bin\'etruy, G. Girardi, R. Grimm and M. M\"uller, \PL{B195}{87}{389};
\\
G. Girardi and R. Grimm, \NP{B292}{87}{181}
\item
\label{Oetal}
B. A. Ovrut and C. Schwiebert, \NP{B321}{89}{163}; \\
B. A. Ovrut and S. K. Rama, \NP{B333}{90}{380}; \PL{B254}{91}{138}
\item
\label{DFKZ1}
J.-P. Derendinger, S. Ferrara, C. Kounnas and F. Zwirner,
\NP{B372}{92}{145}
\item
\label{CFV}
S. Cecotti, S. Ferrara and M. Villasante, Int. J. Mod. Phys.
\underline{A2} (1987) 1839
\item
\label{BGG}
P. Bin\'etruy, G. Girardi and R. Grimm, \PL{B265}{91}{111}
\item
\label{ABGG}
P. Adamietz, P. Bin\'etruy, G. Girardi and R. Grimm, \NP{B401}{93}{257}
\item
\label{W}
E. Witten, \PL{B155}{85}{151}
\item
\label{DFKZ2}
J.-P. Derendinger, S. Ferrara, C. Kounnas and F. Zwirner,
\PL{B271}{91}{307}
\item
\label{DRSW}
M. Dine, R. Rohm, N. Seiberg and E. Witten, \PL{B156}{85}{55}
\item
\label{FILQ}
A. Font, L. E. Ib\'a\~nez, D. L\"ust and F. Quevedo, \PL{B245}{90}{401};
\\
S. Ferrara, N. Magnoli, T. R. Taylor and G. Veneziano, \PL{B245}{90}
{409}; \\
H.P. Nilles and M. Olechowski, \PL{B248}{90}{268};\\
P. Bin\'etruy and M.K. Gaillard, \PL{B253}{91}{119}
\item
\label{duality}
K. Kikkawa and M. Yamasaki, \PL{149B}{84}{357}; \\
N. Sakai and I. Senda, Progr. Theor. Phys. \und{75} (1986) 692
\item
\label{KU}
T. Kugo and S. Uehara, \NP{B226}{83}{49};
\item
\label{CFGVP}
E. Cremmer, S. Ferrara, L. Girardello and A. Van Proeyen, \PL{116B}{82}
{231}
and \NP{B212}{83}{413}
\item
\label{GS}
M. B. Green and J. H. Schwarz, \PL{149B}{84}{117}

\item
\label{WB}
J. Wess and J. Bagger, {\it Supersymmetry and Supergravity,}
2nd edition. Princeton University Press (1992)

\item
\label{FGKVP}
S. Ferrara, L. Girardello, T. Kugo and A. Van Proeyen,
\NP{B223}{83}{191}
\item
\label{KTPVN}
M. Kaku, P.K. Townsend and P. van Nieuwenhuizen,
\PR{D17}{78}{3179}
\item
\label{KU2}
T. Kugo and S. Uehara, \NP{B222}{83}{125}
\item
\label{PvN}
P. van Nieuwenhuizen, Phys. Rep. \underline{68} (1981) 189;
section 4.4, equation (16)


\item
\label{CO}
G. L. Cardoso and B. A. Ovrut, \NP{B369}{92}{351}
\item
\label{CO2}
G. L. Cardoso and B. A. Ovrut, \NP{B392}{93}{315}
\item
\label{DS1}
M. Dine and N. Seiberg, \PRL{55}{85}{366}
\item
\label{BFQ}
C. Burgess, A. Font and F Quevedo, \NP{272}{86}{661}
\item
\label{DS}
M. Dine and N. Seiberg, \PRL{57}{86}{2625}
\item
\label{J}
D.R.T. Jones, \PL{B123}{83}{45}
\item
\label{SV}
M.A. Shifman and A.I. Vainshtein, \NP {B359}{91}{571};
see also \NP{B277}{86}{456}

\item
\label{GT}
M. K. Gaillard and T. R. Taylor, \NP{B381}{92}{577}
\item
\label{AKMRV}
For a review, see D. Amati, K. Konishi, Y. Meurice,
G. C. Rossi and G. Veneziano, Phys. Rep. \underline{162} (1988) 169
\item
\label{DIN1}
J.-P. Derendinger, L. E. Ib\'a\~nez and H. P. Nilles, \PL{B155}{85}{65}
\item
\label{FGN}
S. Ferrara, L. Girardello and H. P. Nilles, \PL{B125}{83}{457}
\item
\label{VY}
G. Veneziano and S. Yankielowicz, \PL{B113}{82}{231};\\
T. R. Taylor, G. Veneziano and S. Yankielowicz, \NP{B218}{83}{493}
\item
\label{MR}
A. de la Macorra and G.G. Ross, \NP{B404}{93}{321}
\end{enumerate}

\normalsize
\end{document}